\documentclass[
superscriptaddress,
nofootinbib,
amsmath,amssymb,
longbibliography,
twocolumn 
]{revtex4-2}

\usepackage{wrapfig}
\usepackage{graphicx}
\usepackage{lineno}
\usepackage[utf8]{inputenc}
\usepackage{rsfso}
\usepackage{amsmath,color}
\usepackage{graphicx}
\usepackage{dcolumn}
\usepackage{bm}
\usepackage{amsfonts}
\usepackage{amsthm}
\usepackage{breqn}
\usepackage{xcolor}
\usepackage{hyperref}
\usepackage{framed}
\usepackage{enumitem}

\setlist[itemize]{leftmargin=0pt}

\usepackage{amsmath}
\usepackage{amssymb}
\usepackage{xcolor}
\usepackage[normalem]{ulem}

\usepackage{tabularx}
\usepackage{booktabs}

\newcommand{\Du}{{\rm Du}}

\newcommand\mc[1]{\mathcal{#1}}

\newcommand\dd{\textnormal{d}}
\newcommand\beq{\begin{equation}}
\newcommand\eeq{\end{equation}}
\newcommand\beqa{\begin{eqnarray}}
\newcommand\eeqa{\end{eqnarray}}

\newcommand\etal{\textit{et al.}}
\newcommand\UU{{\cal U}}

\newcommand\tn[1]{\textnormal{#1}}

\newcommand{\Pe}{\textnormal{Pe}}

\newcommand{\be}{\begin{equation}}
\newcommand{\ee}{\end{equation}}

\def\exp{\textnormal{exp}}

\def\A{\mc{A}}
\def\Du{\tn{Du} }

\begin{document}

\title{Diffusio-osmotic transport in nanochannels}

	\author{Lydéric Bocquet}
    \email{lyderic.bocquet@ens.fr}
    \affiliation{\rm Laboratoire de Physique de l'Ecole Normale Supérieure\\ CNRS and ENS \\ \rm 24 rue Lhomond 75005 Paris}


\date{\today}


%
%
%

\begin{abstract}
\vspace{-0.5\baselineskip}
In this chapter, I will enter into the roots of entropically-driven transport with a focus on diffusio-osmotic transport in nanochannels. 
Diffusio-osmosis is  a subtle surface transport, originating  in  entropic driving forces occuring within the diffuse layers at solid boundaries. 
Specifying diffusio-osmosis to nanochannels may first look like a marginal refinement, yet it reveals that 
osmotic drivings can arise in channels and membranes without the prerequisite of semi-permeability, so that diffusio-osmosis extends the domain of existence of entropically driven transport.  
Osmosis and diffusio-osmosis are two faces of the same phenomenon, naturally embedded in an Onsager framework and quantified by local and global force balances. This perspective clarifies why nanochannels are privileged arenas where diffusio-osmosis and its consequence do flourish. Throughout the chapter, I discuss a set of conceptually relevant examples to show how diffusio-osmosis ``pops up''  in various situations: as enhanced diffusion, mechano-sensitivity, rectified osmotic flows and, ultimately, as a lever for osmotic energy conversion from single nanopores to membrane modules approaching industrial reality.

\end{abstract}
\maketitle

This chapter is part of the book {\it Diffusiophoresis and Diffusioosmosis: Theory, Experiment and Applications}, Editors Ankur Gupta, Guido Bolognesi,   Soft Matter Series of the Royal Society of Chemistry (DOI:10.1039/9781837678136).

\tableofcontents      

\section{Introduction to entropic forces}

\subsection{General perspective}
Diffusio-osmosis refers to the flow of a liquid on a solid boundary, induced by the gradient of a solute along the interface.
This concept broadens the traditional idea of osmosis by recognizing that entropic forces can also act within the interfacial layers at solid surface boundaries, and not only across semi-permeable membranes. 
From this perspective, the phenomenon of diffusio-osmosis might more appropriately be termed {\it epi-osmosis},  highlighting its interfacial roots.

Now, is there any interest -- as suggested by the title of this chapter -- to specify the exploration of diffusio-osmosis 
to {\it nanochannels}. The terminlology ``nanochannels'' suggest a strong spatial confinement, but confinement is not a prerequisite to diffusio-osmosis, only interfaces. Actually, when confinement is very strong, the solute will be (partly or totally) rejected from the channel and lead to bare osmosis. Accordingly the very notion of diffusio-osmosis as a surface induced transport looses its meaning and it gets confused with bare osmosis. Using the description to be discussed below, the interpretation of diffusio-osmosis is crystal clear  when the thickness of the interfacial layers is small compared to the confinement. But osmosis and diffusio-osmosis mechanisms become entangled for thick interfacial layers. 

But a ``reverse'' point of view reveals a most interesting aspect of diffusio-osmosis in nanochannels. Indeed, thanks to diffusio-osmotic transport, osmotic drivings can occur in channels and membranes without the prerequisite of semi-permeability ({\it i.e.} when the solute is rejected from the channel), suggesting that osmotic transport can occur even in non-selective pores. Therefore 
surface-induced osmosis extends the conditions of applicability of 
standard osmosis to less constrained conditions.  This is a considerable asset, and it has strong consequences on various applications, allowing revisiting concepts such as osmotic energy or desalination. 

{\it Disclosure --} 
This chapter is {\it not} a review of diffusio-osmosis in nanochannels. They are actually several good reviews about osmotic phenomena, with some emphasis on diffusio-osmosis: starting for example with the old but excellent review on diffusiophoresis by Anderson \cite{anderson1989colloid}, the review by Marbach and myself on osmosis \cite{marbach2019}, the review by Shim  
\cite{shim2022diffusiophoresis}, and of course the other chapters of this book.
The litterature has been discussed exchaustively in these reviews, in particular the experimental one, and I will not repeat it here. 
This chapter is merely intended as an open discussion on osmotic forces and electrokinetic phenomena, emphasizing aspects that I find particularly insightful rather than aiming for completeness. I will delve into the underlying physical mechanisms governing diffusio-osmosis, their manifestation in nanoscale transport, and selected illustrative examples.
On some aspects, I have chosen to examine the transport mechanisms and their couplings in great technical detail, clarifying how they can be described and modeled. Accordingly, I will  discuss the subtle difference between osmotic and diffusio-osmotic transport and its implication. Also, 
for the case of salt as a solute, I will explore in great details the two limits of thin and thick Debye layers. 
This approach aims to facilitate the analysis and interpretation of experimental data and theoretical predictions while identifying the distinctive signatures of osmotic forces. 

Ultimately, I just want these notes to support readers who wish to engage deeply with this captivating topic.
This will be the guiding thread of this chapter.

%


\subsection{A brief chronology}

The concept of diffusio-osmosis itself -- as well as its reciprocal effect, the excess solute flux under flow -- has been first introduced in a seminal paper of Derjaguin {\it et al.} in 1947 (re-published in english in \cite{derjaguin1947}). 
The notion that solute (or temperature) gradients at interfaces can induce liquid motion closely parallels Marangoni flows, which arise along the interface between two phases due to gradients in surface tension \cite{anderson1989colloid,bacchin2019interfacially}.
However, the Marangoni effect is expected to vanish on a solid surface as the viscosity of the solid phase diverges \cite{anderson1989colloid}.

The key remark of Derjaguin \etal ~ is then to point out the importance of the 
``{\it diffuseness of adsorption layers}''. 
Quoting Derjaguin \etal, the diffuse nature of the interface implies``{\it that a considerable part of the adsorption of the
boundary layer is movable, similarly to the situation for a double layer of ions.
Hence, it is natural to make a conclusion that kinetic phenomena must exist in the
solutions of neutral molecules. }''

Diffusio-osmosis is therefore rooted in the extended width of the interfacial structure close to solid surface. This structure is for example the electric double layer (EDL) for electrolytes on charged surfaces, or an adsorption or depletion layer for neutral solute.
As I will describe below, an osmotic pressure gradient is built within the diffuse layer under a concentration gradient, leading to fluid flow along the surface. 

Strangely enough, the phenomenon has  been quite ignored after Derjaguin \etal ~article and there was little interest in the concept of diffusio-osmosis at a surface. Interest rose again in the 1980's in the context of colloidal diffusiophoresis with the works of Prieve, Anderson among others, as quoted in  \cite{anderson1989colloid}. It revided again in the 2000's with the introduction of microfluidic tools, which allow an exquisite control of chemical gradients, see {\it e.g.} Abecassis \etal \cite{abecassis2008boosting,abecassis2009osmotic} and Palacci \etal \cite{palacci2012osmotic}.

The emergence of nanofluidics since the 2010's allowed further progress in the exploration of the diffusio-osmotic transport at nanoscales. The possibility of building individual and well-controlled nanochannels allowed exploring fundamentally the fluid transport properties under various driving, including chemical gradients \cite{bocquet2020nanofluidics,kavokine2021fluids}. Nanoscale channels can be fabricated in various dimensions: from {\it 0-D} pores across 2D membranes, to {\it 1-D} nanotubes made of various materials (graphite, boron-nitride), to {\it 2-D} channels fabricated by van der Waals assembly \cite{kavokine2021fluids}. 
Experiments have highlighted a whole cabinet of curiosities of transport phenomena, many of them involving diffusio-osmotic transport, {\it e.g.} the measurement of giant ionic currents under salinity gradients across boron-nitride nanotubes, single layer $MoS_2$, or 2D activated carbon channels \cite{siria2013giant,feng2016single,emmerich2022enhanced}. 

Note however, there has been little direct experimental  demonstration of diffusio-osmosis. One exception is the measurement of fluid flow in nanochannels ($\sim 150$nm) under the gradients of various solute species \cite{lee2014osmotic,lee2017nanoscale}.

\section{The Onsager transport matrix}

In full generality, let us consider the transport of an electrolyte and/or a neutral solute across a porous material (possibly constituted of a collection of nanochannels). 
{In full generality, a linear relation can be written between thermodynamic forces and fluxes  \cite{degroot2013non} and one can accordingly relate 
 the solvent flux $Q_w$, excess solute flux $J_s-c_{\rm sol} Q_w$ and electric current $I_e$ to the hydrodynamic pressure drop $\Delta p$, chemical potential drop $\Delta \mu_{rm sol}$ and the applied electric potential drop
$\Delta V$, as a linear relationship
\begin{equation}
\label{Larray2}
\left(\begin{array}{c} Q_w \\ {J_s-c_{\rm sol}Q_w} \\ I_e \end{array}\right)= 
{\mathbb L}
\times \left(\begin{array}{c} - {\Delta p} \\ - {\Delta \mu_{\rm sol} } \\ - {\Delta V } \end{array}  \right),
\end{equation}
where the transport matrix is symetric and positive definite according to the Onsager principle \cite{degroot2013non}. The interpretation of each term of this matrix is 
reminded in Fig.\ref{fig:TransportMat}. 
The diagonal terms correspond to the standard hydrodynamic permeability, diffusion and electric conductance. The off-diagonal terms correspond to cross effects. 

\begin{figure}[h!]
\centering
  \includegraphics[width=0.5\textwidth]{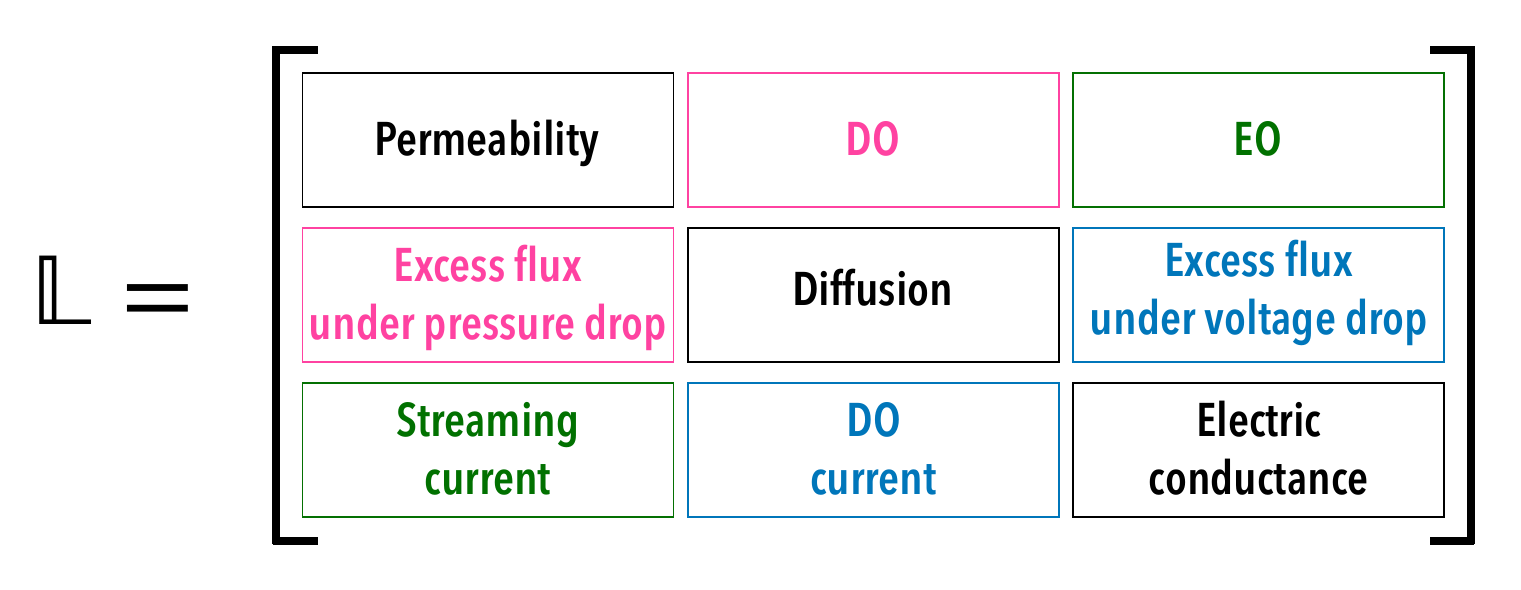}
  \caption{{\textbf{: Transport Matrix.} Explicit transport matrix ${\mathbb L}$ as introduced in Eq.\eqref{Larray2}, with explanation for each term and colors highlighting the Onsager symmetry. Diagonal terms are associated  with permeability (characterizing solvent flux under a pressure drop), diffusion (characterizing solute flux under an applied solute gradient) and electrical conductance (characterizing ionic current under an applied electric field), respectively. The off-diagonal terms correspond to cross effects, with similar mobilities from symmetric terms: diffusio-osmosis (DO) and excess flux under pressure drop; electro-osmosis and streaming currents; diffusio-osmotic currents and excess flux under voltage drop.}}
  \label{fig:TransportMat}
\end{figure}
It is possible to extend this description to include thermal effect, in particular thermo-osmosis \cite{herrero2022fast,ganti2017molecular}.

The (combined) osmotic and diffusio-osmotic (DO) term of interest in this chapter appears in the term relating the flow $Q_w$ to the chemical gradient $\nabla \mu$. 
But due to the Onsager symmetry, they are also associated with the excess flux of solute under pressure drop. 

Following Kedem and Katchasky\cite{kedem1961physical} (and in the absence of electric drivings), it is usual to rewrite these equations in a more explicit form as~
\begin{eqnarray}
&Q= -\mathcal{L}_{hyd} \left( \Delta p - \sigma k_BT \Delta c_{\rm sol}\right), \label{kk1}\\
&J_s= - \mathcal{L}_{D} P \Delta c_{\rm sol} + c_{\rm sol} (1-\sigma) Q, \label{kk2}
\end{eqnarray}
where a small solute concentration is assumed. 
Here $\mathcal{L}_{hyd}$ is the hydrodynamic permeance across the porous material, $\mathcal{L}_{D}={\cal A} D/L$  the solute permeance with $D$ the diffusion coefficient of the solute ($L$ the thickness
of the porous system, ${\cal A}$ the cross section) and  $\sigma$ is a rejection coefficient, $P$ a solute permeability. 
}

\section{Osmosis, with or without a membrane}


\subsection{Back to basics: entropic forces}

Let us start digging into the physics principles of osmosis. Osmosis has a long, sometimes heckled, history: this goes back to the observations of l'Abb\'e Nollet in 1748, to Dutrochet interpretation of endo- and exo- osmosis in 1827, then Pfeffer well-controlled experiment in 1890, to the rationalization of the osmotic pressure by the celebrated van 't Hoff equation in 1892; see for example the recent perspective by Manning and Kay \cite{manning2023physical}, who also point to various misconceptions of the concept of osmosis in the literature.

Osmosis is usually introduced by considering two volumes of a liquid, say water, with different solute concentrations dissolved in it and separated by a semi-permeable membrane. Semi-permeable meaning that the membrane is permeable to water but not to the solute. 
The water being in thermal equilibrium across the membrane, its chemical potential should be equal in the two compartments. Since the solute concentrations differs, this cannot be realized for the same pressure and a difference of pressure builds up between the two reservoirs. For dilute solutes, this reduces to the celebrated van 't Hoff law \cite{marbach2019}:
\beq \Delta \Pi=k_BT \Delta c_{\rm sol}.
\label{vantHoff}
\eeq
with $\Delta c_{\rm sol}$ the solute concentration drop between reservoirs.
In the absence of a piston imposing the  pressure difference, a flux of water $Q_w$ will cross the membrane to equilibrate the two solutions by diluting the reservoir with highest concentration. 
\beq
Q_w={\cal L}_w (\Delta \Pi-\Delta p_h)
\label{Qw}
\eeq
with ${\cal L}_w$ the hydrodynamic permeability of the semi-permeable membrane and $\Delta p_h$ a supplementary hydrostatic pressure difference. The osmotic contribution to the flow is directed from the low to the high concentration reservoir. 

For larger salt concentration, the ideal Van 't Hoff law in Eq.(\ref{vantHoff}) has to be modified to account for the activity of the salt. An osmotic correction factor is introduced as $\Delta \Pi=k_BT \Phi\,\Delta c_{\rm sol}$, the value of which is tabulated in the literature, {\it cf. e.g.} 
the work by Pitzer et al. \cite{pitzer1973thermodynamicsI,pitzer1973thermodynamicsII}.

A first comment is that the resulting osmotic pressure is huge: a difference of solute concentration of 1 M yields an osmotic pressure of 50 bar, the pressure found at -500m below the level of the sea. 
A second comment is that the microscopic properties of the membrane disappear in the van 't Hoff formula. This is somewhat counterintuitive as the membrane selectivity is at the core of the osmotic process.

\subsubsection{The membrane as an insurmountable energy barrier}

This suggests actually an alternative perspective on the physical mechanisms of osmosis, highlighting the mechanical origins of the van 't Hoff law. 
The argument was actually presented by Debye in 1923 \cite{debye1923theorie} (see Refs. \cite{manning2023physical} and \cite{marbach2019} for further discussions), but largely forgotten.
It goes as follows:
Since the membrane details disappear in the final formula, one may replace the membrane by a external potential ${\cal U}(x)$ acting on the solute but not on the water molecules. Semi-permeability is ensured by assuming that the energy barrier is large compared to $k_BT$. This simplistic model obviously forgets about any  interaction heterogeneities inside the pores, as well as the pore structure itself. It has mainly an educational value.

Since the maximum energy is large compared to $k_BT$ the solute does not cross the barrier and one considers the solute concentration in both reservoirs $c_{R/L}$ independently.  
At equilibrium, the solute profile takes an equilibrium Boltzmann expression on each (right $R$ or left $L$) side of the membrane :
\beq c_{R/L}=c_{R/L}^\infty\, \exp[-{\cal U}(x)/k_BT].\eeq
The salt concentration difference between the two reservoirs is defined as $\Delta c_{\rm sol}=c_{R}^\infty-c_{L}^\infty$.

Now the force acting on the global fluid reduces to the force acting on the solute, and we obtain (for an area $\cal A$ of the membrane)
\beq 
{\cal F}_{R/L}=\sum_{solute}(-\partial_x {\cal U})=\A \int_0^\infty c_{R/L}(x)(-\partial_x {\cal U})\dd x
\eeq
where the sum runs on the solute molecules in the left or right reservoir. 
Now using the Boltzmann expression for $c_{R/L}(x)$, one gets that $c_{R/L}(x)(-\partial_x {\cal U})=k_BT \partial_x c_{R/L}(x)$. 
Accordingly the integral above leads immediately to the expression for the force due to the membrane and acting on the fluid as
\beq \Delta \Pi\equiv {{\cal F}_{R}-{\cal F}_{L}\over {\cal A}} =k_BT (c_{R}^\infty-c_{L}^\infty).\eeq
It reduces to the van't Hoff expression for the osmotic pressure.


One therefore learns from this simplistic description that {\it the osmotic pressure is essentially the force exerted by the membrane to prevent the solute from passing through it}. 

\subsubsection{A  membrane with partial rejection of the solute}
This description can be extended in many ways. But an immediate and illustrative extension is to consider partially permeable membranes.
In the context of the energy barrier description, this accounts to assuming that the energy barrier $\UU(x)$ is finite so that a finite solute flux crossing the membrane will occur  \cite{manning1968binary,marbach2017osmotic}. Let us assume that $\UU(x)$ is non-zero only over a finite width 
$[{-L/2,L/2}]$, with $L$ the thickness of the membrane. 

In the presence of a {\it finite} energy barrier $\UU(x)$, the concentration profile $c_{\rm sol}(x)$ will obey the Smoluchowski equation:
\begin{align}
0={\partial_t} c_{\rm sol} =& -\partial_x j_s  \nonumber\\
=& -\partial_x\left( -D \partial_x c_{\rm sol} + \lambda c_{\rm sol} \,(-\partial_x \UU) \right),
\label{Smolu}
\end{align}
where $D$ is the diffusion coefficient,  $\lambda=D/k_BT$ the mobility. Note that for simplicity in this section, we do neglect convection 
in Eq.~\eqref{Smolu},  
This convective term is in most cases small for flows versus the diffusive term in nanoporous structures (small P\'eclet limit);
convection effects will be discussed more extensively in the sections below. 
The Smoluchowski equation, Eq.(\ref{Smolu}), can then be easily solved to obtain the expression for the concentration profile $c_{\rm sol}(x)$. 

Now the interactions between the solute and the barrier will  induce a force on the fluid, hence a fluid flow. The fluid velocity $v_w$ obeys the Stokes equation
\begin{equation}\label{Stokes}
\eta\nabla^2v_w- \nabla p + c_{\rm sol}\,(-\partial_x \UU) =0 ,
\end{equation}
The membrane acts on the fluid via a global force on the solute writing  ${\cal F}={\cal A} \int_{-\infty}^\infty c_{\rm sol}(x)(-\partial_x {\cal U})\,\dd x$.
Along the same lines as above, this force can be integrated easily to identify with an apparent osmotic pressure now taking the form
\begin{equation}
\Delta \Pi_{\rm app}={{\cal F}\over \A}=\sigma_O \Delta \Pi
\label{Starling}
\end{equation}
where $\Delta \Pi=k_BT (c_{R}^\infty-c_{L}^\infty)$ is the van 't Hoff expression, and the force now introduces a rejection (or Staverman) coefficient $\sigma_O$, which in the present description writes
\begin{equation}
\sigma_O= 1-{1 \over \langle \exp[\beta \UU]\rangle}.
\label{sigmaO}
\end{equation}
with $\langle \exp[\beta \UU]\rangle ={1\over L} \int_{-L/2}^{L/2} dx^\prime\, \exp[+\beta\UU(x^\prime)]$; $\beta=1/k_BT$.
This equation points immediately to the interesting limits of a perfectly semi-permeable membrane (large energy barrier), for which $\sigma_O = 1$, while a perfectly permeable membrane (vanishing energy barrier $\UU=0$) yields $\sigma_O =0$ resulting in no osmotic pressure. 

There is a intrinsic relationship between the rejection coefficient $\sigma_O$ and the solute permeability $P$, defined as $j_s = -D\,P\, {\Delta c_{\rm sol}/L}$.
Indeed integrating the flux, as introduced in Eq.(\ref{Smolu}), one obtains. 
$j_s = -{D\over L} \Delta c_{\rm sol} + {D\over k_BT} {1\over L} \int_{-L/2}^{L/2}  dx\,  c \,(-\UU)$. Since according to Eq.(\ref{Starling}) the last term identifies to 
$\sigma_0 k_BT \Delta c_{\rm sol}$, one deduces that the solute permeability is
\beq
P=1-\sigma_O
\label{Psigma0}
\eeq

In general, one may expect the solute permeability and the rejection to be dominated by the maximum energy barrier, so that $P=1-\sigma_O \sim \exp[-\beta \UU_{\rm max}]$. 

To conclude on this description, we emphasize that this mechanical perspective is mainly pedagogical and serves as a simplistic illustration. But it contains a lot of physics insights on the osmotic phenomenon.
The description can be also easily extended to a broad variety of situations, in particular the effects of finite convection  \cite{manning1968binary}, large solute concentrations \cite{marbach2017osmotic,yoshida2017osmotic}, time-dependent energy barriers and resonant osmosis \cite{marbach2020resonant}, and even non-linear response and osmotic diodes \cite{picallo2013nanofluidic}.  
It also provides the basic physical picture for diffusio-osmotic transport, which we now discuss. 


\subsection{Diffusio-osmosis and osmotic forces at solid interfaces}

This mechanical perspective allows  capturing similarly the phenomenon of diffusio-osmosis and  the physical mechanisms at its origin. 
We now consider a liquid-solid interface along which a gradient of a dissolved solute exists, say $\nabla c_{\rm sol}$ along the direction $x$. The solute is supposed to interact with the surface with some interaction, say $\UU(z)$ as above, but now {\it perpendicular}
to the gradient direction.

\begin{figure}[h!]
    \centering
    \includegraphics[width=0.5\textwidth]{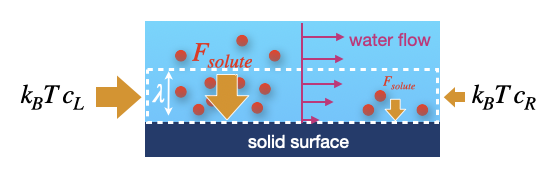}
    \caption{Force balance on diffusio-osmosis. A gradient of solute concentration along a solid surface induces an osmotic pressure drop within the diffuse layer {\it via} the interaction with the solid surface. This induces a diffusio-osmotic water flow along the surface. }
    \label{fig:DO-force}
\end{figure}

Let us start with a back-of-the-enveloppe calculation following the same lines as in the previous reasoning. Under the interaction with the solid, the solute is spread over a diffuse layer within the liquid, say of thickness $\lambda$ (range of  $\UU(z)$). 
The solid will accordingly exert a force on the solute $f_z=\sum_{solute}(-\partial_z U)$ (per unit surface) on this diffuse layer, now in the $z$ direction. Let us emphasize that this force is localized within the diffuse layer and no forces is exerted outside of the layer. Along the same arguments as above, this force will take the form of an osmotic pressure $f_z\sim k_BT c_{\rm sol}(x)$, which is transmitted as an isotropic pressure to the fluid, but only within the diffuse layer. The concentration gradients yields therefore a pressure gradient on the liquid.  The force balance between the osmotic pressure within the diffuse layer and the hydrodynamic friction on the solid surface,  as sketched on Fig.\ref{fig:DO-force}, thus writes
\beq
\eta {v_{DO}\over \lambda} \times L \approx -\lambda \times k_BT \Delta c_{\rm sol}
\label{VDOqualitatif}
\eeq
with $v_{DO}$ the fluid hydrodynamic velocity and $\Delta c_{\rm sol}=c_R-c_L$ the difference of solute concentration over a distance $L\gg \lambda$ and far from the solid boundary; $\Delta c_{\rm sol}=L\times \nabla c_{\rm sol}$.
We deduce that the DO velocity should scale as
\beq
 {v_{DO}}  \approx {k_BT\over \eta} \lambda^2 \times  (-\nabla c_{\rm sol})
\eeq
The lengthscale $\lambda$ is the range of the solute-solid interaction, hence typically in the order of the molecular scale for the solute. I will come back on the sign of $v_{DO}$ below. 

Now, for an electrolyte, the diffuse layer identifies with the Debye layer and $\lambda^{-2}=8\pi \ell_B c_s$, with $\ell_B=e^2/4\pi\epsilon k_BT$ the Bjerrum length ($\epsilon$ the dielectric constant), and one imposes a gradient of salt concentration $c_s$.  Note that the concentration of ions as a solute is $c_{\rm sol}=c_++c_-=2 c_s$ (accounting for both anion and cation contributions). Accordingly for electrolytes, 
\beq
 {v_{DO}}  \approx {k_BT\over 8\pi \eta\ell_B}  \times  2\times (-\nabla \log c_s)
\eeq
where $ (-\nabla c_s)$ is the salt concentration gradient.

In this case, the diffusio-osmotic mobility for a charged solute takes the approximate expression
\beq 
D_{DO} \approx {k_BT\over 4\pi \eta\ell_B} 
\label{DDOapp}
\eeq 
$D_{DO}$ has the units and the order of magnitude of a molecular diffusion coefficient. Hence, one typically expects  $D_{DO}\sim 10^{-9}$m$^2$/s. Accordingly,  a gradient occuring over a length of $1\mu$m to fix ideas
({\it i.e.} $\nabla \log c_\infty \sim (\mu{\rm m})^{-1}$) generates a flow in the millimeter per second range.\\


Let us now make the argument more precise. This reproduces the classical derivation as developed in \cite{anderson1989colloid,marbach2019} and other chapters of this book. 

\subsubsection{neutral solutes}
The relaxation towards equilibrium within the nanometric diffuse layer occurs very fast, typically over a timescale $\tau_\lambda \sim \lambda^2/D$, with $\lambda \sim $ nm the width of the diffuse layer, so that $\tau_\lambda \sim$ ns.
 Under an external potential $\UU$, the $z$-dependent profile of the solute (in the diffuse layer perpendicular to the surface)  then quickly relaxes to a local Boltzmann equilibrium $c(x,z)=c_{\rm sol}(x)\, \exp[-{\cal U}(z)/k_BT]$, with $c_{\rm sol}(x)$ the solute concentration far from the surface. 
 
 The force balance in the $z$ direction writes $0=-\partial_z p + c(x,z)(-\partial_z \UU)$. Therefore, 
the force per unit surface acting on the solute/liquid due to the solid interface (over a semi-infinite space $[z,\infty]$) rewrites in terms of a local and isotropic osmotic pressure
\beq 
f_z(x)=  \int_z^\infty c(x,z^\prime)(-\partial_z \UU)\dd z^\prime= k_BT \left[c(x,z)-c_{\rm sol}(x)\right]
\eeq
Outside of the diffuse layer ($z\rightarrow \infty$), $c(x,z)\simeq c_{\rm sol}(x)$ and the force density $f_z\simeq 0$ as expected. 
The corresponding hydrostatic pressure within the diffuse layer thus writes 
\beq
p=p_\infty +k_BT (c(x,z)-c_{\rm sol}(x)).
\label{hydrostat}
\eeq

The pressure gradient induces a shear flow along $x$  obeying Stokes equations
\begin{eqnarray}
&\eta {\partial^2 v_x\over \partial z^2}&= \nabla_x \left( k_BT \left[c(x,z)-c_{\rm sol}(x)\right]\right)\nonumber \\
&&= k_BT \nabla_x c_{\rm sol} \left( e^{-\beta \UU(z)}-1\right)
\end{eqnarray}
and  the DO velocity is finally obtained as
\beq
v_{DO}={k_BT\over \eta} \lambda^2 (-\nabla_x c_{\rm sol})
\eeq
with ``$\lambda^2$'' defined as
\beq
\lambda^2= \int_0^\infty z \left[\exp[-\beta\UU]-1\right]\,dz
\eeq
Formally, ``$\lambda^2$'' is put under guillemet since its sign depends on the details of the solute interaction with the wall: $\lambda^2$  is indeed positive for an attractive wall-solute interaction, but can reverse sign for repulsive walls.
For example for a neutral polymer repelled from a surface via excluded volume effects, one expects $\lambda^2 \simeq - R_g^2<0$, with $R_g$ the radius of gyration of the polymer \cite{lee2014osmotic}.

Note finally that these results can be generalized at high salt concentration \cite{marbach2017osmotic}
\begin{equation}
v_\infty= K_{DO} (-\partial_x \Pi[c_{\rm sol}(x)]),
\label{vDO}
\end{equation}
with $\Pi(c)$ the general expression for the osmotic pressure beyond the ideal solution limit, and the diffusio-osmotic mobility $K_{DO}$ is now given as
\begin{equation}
K_{DO}=  \frac{1}{\eta}\int_0^\infty dz^\prime\, z^\prime\, \left({c(x,z^\prime)\over c_{\rm sol}(x)} -1\right).
\label{KDO}
\end{equation}

This emphasizes that the driving force is indeed the osmotic pressure. 

%

\subsubsection{charged solutes}
For charged solutes the derivation proceeds along the same lines \cite{prieve1984motion}. 
The pressure profile follows from the local force balance as discussed above and yields an expression similar
to Eq.(\ref{hydrostat})
\begin{equation}
p(x,z)=p_\infty +  2 k_BT\, c_s(x) \left[ \cosh \psi(z) -1\right] 
\end{equation}
with $\psi(z)=eV(z)/k_T$ the (dimensionless) electric potential profile, obeying the Poisson-Boltzmann equation; see Appendix.
Let us remind that here  $c_s$ is the salt concentration, so that the total ion concentration is $c_{\rm sol}=2 c_s$.

The velocity profile $v_x(z)$ stems from the pressure gradient $\nabla_x p$ and writes
\begin{equation}
\eta {d^2 v_x \over dz^2} =  2\, k_BT\,  \left[ \cosh \psi(z) -1\right] \times {\nabla_x c_s} 
\label{Vx}
\end{equation}

Since the osmotic pressure gradient is localized within the EDL layer only, this leads to plug like flow beyond the EDL with
a DO velocity given by
\beq
v_{DO}= D_{DO} (-\nabla \log c_s)
\eeq
where the mobility $D_{DO}$ now takes the general expression
\begin{equation}
D_{DO}= \frac{k_BT }{\eta}\,  \int_0^\infty dz\, z\, \left[{c(z)} -2c_s\right].
\label{DDO-general}
\end{equation}
with $c(z)=c_+(z)+c_-(z)$ the total ion concentration.
As a side note it is interesting to compare this result to that for electro-osmotic mobility:
\begin{equation}
\mu_{EO}= \frac{1 }{\eta}\,  \int_0^\infty dz\, z\, {n_c(z)} 
\label{EO-general}
\end{equation}
with $n_c(z)=-e(c_+(z)-c_-(z))$ the charge density, leading to the Helmholtz-Smoluchowski expression $\mu_{EO}=-{\epsilon \zeta\over \eta}$, with $\zeta$ the zeta potential of the surface.

Within the PB framework, the DO mobility relates to the electrostatic potential as
 $c(z)-2c_s=2 c_s\left[ \cosh \psi(z) -1\right]$. 
The mobility then takes the expression \cite{anderson1989colloid}
\begin{equation}
D_{DO}= { k_BT\over 2\pi \eta \ell_B} 
\times \log \left[\cosh^2{\psi_0\over 4} \right]
\label{DDO}
\end{equation}
with $\psi_0=eV_0/k_BT$ is the dimensionless surface potential $V_0$.
It can be equivalently expressed in terms of the surface charge $\sigma=-e \Sigma$, since $ \cosh^2{\psi_0\over 4} 
= (1-\gamma^2)^{-1}$, with $\gamma=\tanh {\phi_0\over 4}$ solution of $\gamma^2+2{\ell_{GC}/\lambda_D} \gamma -1=0$;
$\ell_{GC}=(2\pi \Sigma \ell_B)^{-1}$ is the Gouy-Chapmann length (see Appendix for the PB solutions).

If ions have different diffusivities ($D_+\ne D_-$), a diffusional electric field induced by the difference of ion mobility builds up, with
\beq
E_{\rm diff}={k_BT\over e} \beta_s \nabla \log c_s
\eeq  
with $\beta_s=(D_+-D_-)/(D_++D_-)$. This a supplementary contribution to the diffusio-osmotic which writes
\beq
D_{DO}^{\rm diff}=\mu_{EO} \times {k_BT\over e} \beta_s, 
\label{DOdiff}
\eeq
with $\mu_{EO}=-{\epsilon \zeta\over \eta}$ and $\zeta$ the zeta potential of the surface. Note that in principle $\zeta=V_0$, but experimentally this identification is not always pertinent, due molecular effects at the surface,  for example variations of the dielectric profile or interfacial slippage effects . 


Typically $\log \left[\cosh^2{\psi_0\over 4} \right]\sim{\cal O}(10^{-1})$ for surface potentials in the range of a few $k_BT/e$ and
the expression in Eq.(\ref{DDO}) qualitative agrees with the rule-of-thumb result in Eq.(\ref{DDOapp}). 
However, 
the full expression in Eq.(\ref{DDO}) depends on the details of the EDL via the surface potential $\Psi_0$ and zeta potential $\zeta$.  The DO mobility may thus depend on the salt concentration via the salt dependence of $\Psi_0$ and $\zeta$.
The mobility depends on the properties of the EDL and the surface charging mechanism ({\it e.g.} fixed surface charge, fixed potential or charge regulation).
Typically, the surface potential usually decays with salt concentration. For example for fixed surface charge and in the Debye-H\"uckel regime, $\Psi_0 \sim 1/c_s^{1/2}$.
Also, as reported by Kirby {\it et al.} \cite{kirby2004zeta}, the zeta potential decays with concentration as $\zeta = - a_1 \log_{10} c^\star$ for a wide range of solute
concentrations (here for symmetric electrolytes with a valence of one; $c^\star$ is the salt concentration expressed in M; typically $a_1 \sim 20$mV).
The DO mobility is therefore expected to (slowly) decrease for high salt concentration. 

%
%
%

\subsubsection{Excess solute flux}
The calculation of the excess solute flux corresponds to the symmetric cross-effect in the Onsager matrix, Eq.(\ref{Larray2}): an excess solute flux under a pressure-driven flow.
The symmetry of the Onsager matrix shows that this excess flux is proportional to the pressure-gradient with 
the same diffusio-osmotic mobility, {\it i.e.}
\beq
J_s-c_{\rm sol} Q= {{\cal A}} \times {D_{DO} \over k_BT}  (-\nabla p)
\label{excessflux}
\eeq
where ${\cal A}$ is the cross section of the considered channel. 

This is easily derived. Let me consider a geometry of a channel confined between two independent but similar surfaces, and height much larger than the
thickness of the diffuse layer. Under a
pressure gradient, a flow is generated in the channel. Since there is an excess of solute (salt) in the diffuse layer,
there is an overall excess solute flux as compared to the bulk one, {\it i.e.} a non-vanishing $J_s-c_{\rm sol} Q$.
Note again that $c_{\rm sol}=2c_s$ for a salt.
The excess solute flux convected by flow takes the expression:
\beq
J_s-c_{\rm sol} Q= \int d{\cal A}\,  \left({c} -c_{\rm sol}\right) \times v_w
\eeq
In the diffuse layer, the Poiseuille flow is approximated as $v_w(z)\simeq \dot\gamma_0\,z$, with $\dot\gamma_0$ the shear rate at the surface and $z$  the coordinate perpendicular to the wall. 
For a Poiseuille flow in parallel slits with height $h$, $\dot\gamma_0= {h\over 2\eta}(-\nabla p)$, so that
\beq
J_s-c_{\rm sol} Q= {\cal A}{1\over \eta} \int_0^\infty dz\, z \left({c} -c_{\rm sol}\right) \times (-\nabla p)
\eeq 
where the contributions of the two walls are added up. 
One recognizes the expression of $D_{DO}$ in Eq.(\ref{DDO-general}), yielding  Eq.(\ref{excessflux}).

\subsubsection{Ion specific effects}
As a complementary remark, it is worth  mentioning that the above approach for the DO transport of electrolytes is universal in the sense
that it does not depend on the specifics of each ion specie. This is of course a crude approximation and ion specific effects
are expected to emerge at interfaces. In a simple description, ion specific effect can be accounted for by simply combining the results
for neutral and charged species. Indeed, in addition to the electrostatic potential, ions close in the diffuse layer will be associated with specific
adsorption free energies and profiles, say ${\cal U}_\pm (z)$. This supplementary free energy may originate in various mechanisms, {\it e.g.}  van der Waals attraction, steric effects, image charges, (ion volume dependent) solvation energy at the surface, etc.
Such effects should be added in the Poisson-Boltzmann description, as done for example in Ref.\cite{huang2007ion}, in the context of EK transport. The ion concentration profiles now obey 
\beq
c_\pm(z)=c_s \exp\left[ \mp \psi(z) - \beta {\cal U}_\pm (z)\right]
\eeq
with $\psi=eV/k_BT$ the dimensionless electrostatic potential, together with  the Poisson equation
\beq
\Delta \psi = -4 \pi \ell_B (c_+(z)-c_-(z)) 
\eeq
hence constituting a modified PB equation. Note that spatial variations of the
dielectric constant may be included as well. The DO mobility, defined in general in Eq.(\ref{DDO-general}), will then
be altered by the ion specificity.
Such effects have been investigated for osmotic transport in classical and {\it ab initio} molecular dynamics in Ref. \cite{lee2017nanoscale,joly2021osmotic}. 
They reveal the subtle influence of molecular effects on interfacially driven transport and in particular diffusio-osmosis.

\subsection{Various add-ons for diffusio-osmosis}

The formulation of diffusio-osmosis outlined in the previous sections can be extended by incorporating various additional elements and  refinements. I discuss some of them in the present sections, altough the list is non exhaustive.

\subsubsection{Hydrodynamics slippage effects}

The classical description of fluid flows at surface usually assume no slip at the interface, so that the fluid velocity at the surface vanishes.
This widely used boundary condition has however no fundamental basis and in general a finite slip velocity may
occur at the solid surface \cite{bocquet2010nanofluidics,kavokine2021fluids,aluru2023fluids}. This is accounted for by the partial slip boundary condition which writes
\beq
b {\partial v_w\over \partial z}{(z=0)} = v_w(z=0)
\eeq
The slip length $b$ is defined in terms of the fluid-solid friction coefficient $\lambda_f$, as
\beq
b={\eta\over \lambda_f}
\eeq
where $\eta$ is the bulk fluid viscosity. While the slip length is merely a convenient quantity to quantify slippage, 
the friction coefficient $\lambda_f$ is a well-defined physical quantity,
expressing the friction force on the solid in terms of the slip velocity as $F_f = -\lambda_f\, {\cal A} v_{slip}$
(${\cal A}$ the contact area). It can be expressed in terms of the molecular dynamics at the surface in the
form of a Green-Kubo relationship \cite{bocquet2010nanofluidics}:
\beq
\lambda_f = {1\over {\cal A} k_BT} \int_0^\infty dt\, \langle F_f(t)\cdot F_f(0)\rangle_{\rm equ}
\eeq
with $F_f$ the fluctuation microscopic force of the fluid acting on the solid surface. The average is quantified in a fluid at equilibrium.
I refer to previous reviews \cite{bocquet2010nanofluidics,kavokine2021fluids,aluru2023fluids} for more exhaustive discussions of the
slip phenomenon and its origin.

Hydrodynamic slippage has been observed on many surfaces, with larger slippage associated with more hydrophobic
materials. Typically the slip length on hydrophobic materials is in the range of tens of nanometers: $b \sim 10-30$nm.
The slip length of water on pristine graphite is in the range of 10 nanometers. However,
the water-carbon interface is very specific, with considerable slippage measured for 
water in carbon nanotubes with slip lengths in the range of hundreds of nanometer to microns, and increasing for smaller radii
\cite{kavokine2021fluids,aluru2023fluids}. This feature was accounted for by the occurence of quantum solid-liquid 
friction on graphitic surfaces \cite{kavokine2022fluctuation}.

It is obvious that hydrodynamic slippage, which occurs within the diffuse layer, does strongly affect the surface-induced transport phenomena. 
This can be immediately recognized in the force balance within the diffuse layer, in Eq.(\ref{VDOqualitatif}),   between the osmotic pressure and  hydrodynamic friction. Within the diffuse layer with width $\lambda$, the flow is now sheared over the length $\lambda +b$ (instead of $\lambda$), reducing accordingly the viscous stress, so that 
\beq
\eta {v_{DO}\over {\lambda+b}} \times L \approx -\lambda \times k_BT \Delta c_{\rm sol}
\label{VDOqualitatif2}
\eeq
with $v_{DO}$ the DO velocity and $\Delta c_{\rm sol}=c_R-c_L$ the difference of concentration over a distance $L\gg \lambda$ and far from the solid boundary. 
This shows that the DO mobility is enhanced by a factor $b/\lambda$ as
\beq
v_{DO}= {k_BT \over \eta} \lambda^2\left( 1+ {b\over \lambda}\right) \times (-\nabla c_{\rm sol})
\eeq
As shown in \citenum{ajdari2006giant}, the enhancement factor $b/\lambda^{\prime}$ is rather calculated in terms of the length $\lambda^{\prime}$ defined as
\beq
\lambda^{\prime}={\int_0^\infty  dz\, z (c(z)-c_{\rm sol})\over {\int_0^\infty dz\, (c(z)-c_{\rm sol})}}
\eeq
As a rule of thumb, in the case of salt as a solute, the length $\lambda^{\prime}$ reflects the structure of the electric double layer -- see appendix for a reminder of the Poisson-Boltzmann description. Accordingly, one may verify that the length  $\lambda^{\prime}$  reduces to the Debye length 
$\lambda_D$ for weakly charged surfaces ($\ell_{GC}\gg \lambda_D$) -- this is the Debye regime. For highly charged surfaces then $\ell_{GC}\ll \lambda_D$, and the length  $\lambda^{\prime} \approx \ell_{GC}$,  the Gouy-Chapmann length. 
These lengths are accordingly in the nanometer (or even sub-nanometer) range, and the enhancement effect for diffusio-osmosis is therefore expected to be considerable. Altogether for salts, 
\beq
D_{DO}=D_{DO}^{\rm no-slip}\times \left( 1 + {b\over \lambda^{\prime}}\right)
\eeq

This expectation has however to be tempered by the fact that the slip length is expected to decrease for charged surfaces, typically as a scaling law $b\propto \Sigma^{-\alpha}$, with the exponent $\alpha$ close to unity, but depending on the molecular charge pattern
\cite{joly2006liquid,xie2020liquid}.

Still, carbon surfaces -- either pristine or ``activated'' -- were shown to exhibit large DO transport which can be explained by a combination of an electrified surface and slippage enhancement \cite{emmerich2022enhanced}. Note that the latter work developped a 
more detailed description of the ion dynamics inside the EDL, involving the individual friction coefficient of the ions on the surface, following the theory developped in \cite{mouterde2019molecular}. Such a refined description is in particular relevant for surface electrification associated with mobile (physisorbed) charges \cite{mouterde2019molecular,asmolov2025enhanced}.
Enhanced DO transport was also reported in Ref. \cite{Lizee2025architecting}, where slippage effects were accounted for to explain a boost of the transport in nanochannel, see Sec. VIII.B.

I finally quote the result on superhydrophobic (SH) surfaces, where slippage effect are massive due to the composite nature of the interface, made of liquid-solid and liquid-vapor contacts. The effective slip length, which results from the combination of no-slip on the solid surface and perfect slip on the vapour interfaces, typically behaves as $b_{\rm eff} \sim a/\phi_s$, with $a$ the characteristic size of the solid tips and 
$\phi_s$ the solid fraction ($\phi_s\rightarrow 0$ for SH surfaces) \cite{ybert2007achieving}. Slip length on SH surfaces can reach  tens of microns! \cite{ou2004laminar} 

Accordingly one may show \cite{huang2008massive} that DO transport is massively enhanced over SH surfaces, with a DO velocity under salinity gradient writing as
\beq
v_{DO}\approx  {b_{\rm eff}\over \eta} \times \Gamma \times (-k_BT \nabla \log c_s)
\eeq
with $\Gamma=\int_0^\infty dz\, (c_++c_--2c_s)$ the surface excess at the liquid-vapour interface. Counterintuitively the effect disappears for EO transport on such composite surface, demonstrating the subtlety of the underlying nanoscale transport on heterogeneous surfaces \cite{huang2008massive}.

\subsubsection{Specifics of diffusio-osmosis in 0D nanopores}

It is interesting to briefly comment on diffusio-osmosis across nanopores, {\it i.e.} orifices drilled in 2D membranes,
say of vanishing or very small thickness as compared to its diameter.

Such a situation is particularly relevant for example in the context of osmotic energy harvesting across
2D membranes, as reviewed in \cite{macha20192d}. The situation, which we previously discussed in the review
\cite{marbach2019}, is both simple and subtle. Subtle because the surface is not translationally invariant
and the simple force balance proposed above does not apply anymore. The geometry of the flow is complex
and the corresponding calculation of the DO mobility is accordingly complex and requires advanced mathematical tools \cite{rankin2019entrance,baldock2025scaling}.
It is however simple because the main qualitative effect of a vanishingly small thickness will be to approximate the concentration gradient as
$\nabla c \approx \Delta c/a$, with $a$ the pore radius. This is because the Stokes equation obeyed by the flow
is laplacian by nature, and hence mainly determined by the smallest dimension of the pore on which the gradients are the largest, hence $\nabla c \approx \Delta c/a$.
Accordingly, the DO water flux through an orifice will write as
\beq
Q_{DO} = \kappa_{DO} (-k_BT \Delta \log c_{\rm sol})
\eeq
with $\kappa_{DO}$ a 2D diffusio-osmotic mobility. Now the complexity of the flow leads to a complex behavior of the mobility $\kappa_{DO}$ as
a function of Debye length (and/or interaction range for neutral solutes). This was investigated theoretically in Refs. \cite{rankin2019entrance,baldock2025scaling} using blate-spheroidal coordinates, showing that multiple scaling behaviors for
the mobility $\kappa_{DO}$ versus the size of the pore emerge, depending on the conditions.

In order to grasp the underlying mechanisms, let me consider the simplest situation where the range of the potential for the solute (neutral or charged) is larger than the pore size. Then the potential inside the nanopore is homogeneous, as well as the excess concentration inside the pore, which can be approximated as
$c_{in}\approx c_{\rm sol} \times \exp[- {\cal U}_{in}/k_BT]$, with ${\cal U}_{in}$ the homogeneous potential inside the pore, and $c_{\rm sol}$ the solute concentration far from the pore. Under a pressure drop $\Delta p$, the flow rate $Q_w$ across the
nanopore obeys the so-called Sampson law, as $Q_w={a^3\over 3\eta} (-\Delta p)$ \cite{kavokine2021fluids}, with $a$ the pore radius.
Accordingly  the excess
solute flux is calculated in this limit as 
\beq
J_{\rm sol}-c_{\rm sol} Q_w\simeq c_{\rm sol} (\alpha-1)\times {a^3\over 3\eta} (-\Delta p)
\eeq
with $\alpha=\exp[- {\cal U}_{in}/k_BT]$. Using Onsager symmetry, one deduces that in this limit the DO mobility behaves as
\beq
\kappa_{DO}\approx c_{\rm sol} (\alpha-1)\times {a^3\over 3\eta}
\eeq
In general however, a wealth of scaling behaviors is obtained and they depart from this scaling behaviour. It depends on the Debye length, the pore radius, the local surface charge at the mouth, etc. as summarized in Ref. \cite{baldock2025scaling}. To my knowledge, this behavior has not been exhaustively studied on the experimental side.


\subsubsection{Simulations of diffusio-osmotic transport}

Molecular dynamics (MD) is a tool of choice to study interfacially driven transport, and in particular
diffusio-osmotic flows in nanochannels. Indeed, as any EK effects, DO transport originates in subtle
couplings within the first few nanometers close to the surface so that the continuum approaches
which are at the basis for the description of DO -- see above -- may fail capturing some aspects
of the transport. MD can then be instrumental  to get insights into the molecular mechanisms
at play for interfacial transport. 

For example, MD simulations have put forward the effect of slippage on the DO mobility \cite{ajdari2006giant},
 or the  effects of specific cation/anion adsorption free energies 
on various surfaces, {\it e.g.} on graphene versus hBN \cite{joly2021osmotic}, etc. These various effects require
to go beyond the continuum description of diffusio-osmotic transport given above. 

One interesting example in this context is the study of DO of 
 poly(ethylene)glycol polymers and ethanol in water mixtures at 
silica surfaces \cite{lee2017nanoscale}. In both cases, experiments demonstrate a flow in the 
direction of low-to-high concentration. This is unexpected since in both cases, the solute species are expected
to adsorb on  silica and DO flow with attractive surfaces should be directed from the high-to-low concentrations
(opposite to bare osmosis). MD simulations then showed that the sign reversal of DO results from 
a higher friction of solute molecules close to the surface, so that the molecules in the first layers
are slower than the fluid velocity, contributing less to the surface transport. Hence, even in the presence of solute adsorption, the surface  behaves 
similarly to a repulsive one. This dynamical contrast does alter the DO mobility, which
can even change sign under the interfacial slow-down of the solute. 

This is an example among others for the influence of molecular effects on DO transport which can
be accounted for by MD simulations. Recent works even went beyond classical MD to study DO
within the framework of ab initio molecular dynamics of water and solutes on various 2D materials, 
such as graphene, hBN, and MoS$_2$\cite{joly2021osmotic,bilichenko2024slip}. 
Among various results, they revealed ion specific adsorption on various surfaces.

One difficulty with the direct simulation of diffusio-osmosis is that one should account for chemical gradients as a thermodynamic force, {\it i.e.} $\nabla \mu$ with $\mu$ the chemical potential. Actually, first simulations of diffusio-osmosis rather investigated the
excess solute flux under pressure drop \cite{ajdari2006giant,lee2017nanoscale}, which allows quantifying
the DO mobility via Onsager symmetry. Alternatively, various mobilities were also calculated in equilibrium 
MD simulations using Green-Kubo relationships  \cite{yoshida2014molecular}.
For example, the DO terms appearing in the transport matrix in Eq.(\ref{Larray2}) can be written as
\begin{equation}
L_{21} = L_{12} = \frac{\cal V}{k_B T} \int_{0}^{\infty} \left\langle \big(J_s - c_{\rm sol} Q_w\big)(t) Q_w(0) \right\rangle dt.
\end{equation}
with $J_s(t)$ the fluctuating solute flux and $Q_w(t)$ the fluctuating water flux, and the average is obtained from {\it equilibrium} simulations;
${\cal V}$ is the volume of the system.

However, non-equilibrium chemical drivings along surfaces or across nanopore can be mimicked by applying a specific 
set of molecular forces to the solute and solvent \cite{yoshida2017osmotic, monet2023unified}.
Different forces are applied to each solute and solvent molecules in the microscopic system, with a net zero force. 
The equivalent chemical potential is defined in terms of the individual forces \cite{yoshida2017osmotic}.
This simulates an osmotically driven plug-flow inside a nanochannel.

\subsection{Coupling various sources of osmotic drivings and rejection mechanisms}

\subsubsection{Bare osmosis versus diffusio-osmosis}

The discussion above pointed out the various osmotic mechanisms: ``classical'' osmosis with
a semi-permeable membrane; diffusio-osmosis in the presence of a solid interface.
In a nanochannel with ``large'' thickness, much larger than the interfacial diffuse layer, 
diffusio-osmosis will be the dominant transport mechanism, while classical osmosis will
come into play when the thickness of the pore is small enough to induce rejection of the solute.

For intermediate thickness, both phenomena may coexist. Let us enter into the details of such a situation,
which is discussed in Ref.  \cite{marbach2017osmotic,marbach2019}.
The question may appear as quite academic, but it is illustrative of how the various osmotic transport modes couple.

To simplify, I consider neutral solutes and  explore the coupled osmosis and diffusio-osmosis
within the interaction potential (mechanical) perspective. 
%
We consider accordingly the fluid+solute transport across a nanochannel, with size $L$ along $x$ (spanning $[-L/2;L/2]$), in which the solute is subjected to 
a potential taking the form
\begin{equation}
\UU(x,z) = \UU_{\infty}(x) + {\UU_0(x,z)},
\label{Usep}
\end{equation}
where $\UU(x,z)$ is non zero only between $-L/2$ and $L/2$,  $\UU_{\infty}(x) $ describes the global energy barrier associated with the membrane, and {$\UU_0(x,z)$} describes the specific interaction with the membrane pore surface, and vanishes 
far from the pore surfaces. 

Under a global concentration difference in the reservoirs, a complex density profile will be built
inside the nanopore. Assuming that the relaxation in the diffuse layer (along the $z$ direction) is fast, one may accordingly obtain
the solute density profile $c_{\rm sol}(x,y)$ in terms of the central density $c_\infty(x)$ at the nanopore center, as
\begin{equation}
\label{eqPotentialNew}
\mu[c_{\rm sol}(x,z)] +  \UU(x,z) = \mu[c_{\infty}(x)] +  \UU_{\infty}(x)
\end{equation}
where $\mu[c]=k_BT \log [c\, {v}_0]$ is the chemical potential of the solute specie (${v}_0$ an irrelevant microscopic volume) and the potential $ \UU_0(x,z)=\UU(x,z)-\UU_{\infty}(x)$ is assumed to be vanishingly small at the pore center.
Following the same line as for the diffusio-osmotic description, the pressure profile in the $z$ direction
then writes 
\begin{equation}
p(x,z) - \Pi[c_{\rm sol}(x,z)] = p_{\infty} - \Pi[c_{\infty}(x)].
\end{equation}
with $\Pi=k_BT c_{\rm sol}$ the osmotic pressure. To obtain this expression, one makes use of the Gibbs-Duhem relationship between pressure and chemical potential as $c_{\rm sol}(-\partial_z \UU)=c_{\rm sol}(\partial_z \mu)=\partial_z \Pi[c_{\rm sol}]$ \cite{marbach2017osmotic}. Accordingly the previous equations remain valid beyond the low density solute discussed here. 

The Stokes equation for the velocity along $x$ can then be written 
\begin{align}
\eta \partial_{z}^2 v_x = &- \partial_x(\Pi[c_{\infty}(x)] - \Pi[c_{\rm sol}(x,z)]) - c_{\rm sol}(x,z)(-\partial_x \UU_0) \nonumber\\
 &  - c_{\rm sol}(x,z)(-\partial_x \UU_{\infty}).
\end{align}
This can be formally separated into a bulk, osmotic, flow and a surface, diffusio-osmotic, contribution. 
The osmotic velocity profile $v_{osm}(z)$ is defined as 
\begin{equation}
\eta \partial_{z}^2 v_{osm} =    - c_\infty(x)(-\partial_x \UU_{\infty}),
\end{equation}
while the interfacial (diffusio-osmotic) contribution $ v_x^{DO}=v_x-v_{osm}$ verifies the equation
\begin{align}
\eta \partial_{z}^2 v_x^{DO}= &- \partial_x(\Pi[c_{\infty}(x)] - \Pi[c_{\rm sol}(x,z)]) - c_{\rm sol}(x,z)(-\partial_x \UU_0) \nonumber\\
 &  - (c_{\rm sol}(x,z)-c_\infty(x))(-\partial_x \UU_{\infty}),
\end{align}
As done above, we can calculate the corresponding averaged flux $Q_{osm}$  in terms of the permeability $k_{\rm hyd}$ of the pore as 
\begin{equation}
Q_{osm}=\frac{k_{\rm hyd}}{\eta} \sigma_O {\Delta \Pi\over L}.
\end{equation}
with 
$\sigma_{\rm O} = 1- { L \left( \int_{-L/2}^{L/2} dx\, \exp[+\beta \UU_\infty(x)] \right)^{-1}}$ the bare osmosis rejection coefficient.

On the other hand, rewriting the interfacially-driven contribution leads to the DO contribution which, after some manipulations, takes the form (for $z\ge \lambda$, with $\lambda$ the size of the diffuse layer)
\begin{equation}
v_x^{DO} (x,z\ge \lambda)= K_{DO} \left( \partial_x \Pi[c_{\infty}(x)]     - c_{\infty}(x) (-\partial_x \UU_{\infty}) \right).
\label{vxODO}
\end{equation}
where the DO mobility is defined here as
\begin{equation}
K_{DO}=  \frac{1}{\eta}\int_0^\infty dz^\prime\, z^\prime\, \left({c_{\rm sol}(x,z^\prime)\over c_\infty(x)} -1\right).
\end{equation}
The corresponding flux can then be calculated as
\begin{equation}
 Q_{DO}  = \frac{1}{\mathcal{A}\, L}\int d\mathcal{A}  \int_{-L/2}^{L/2} dx  \, \delta v_x (z) 
=  {K}_{DO} (1-\sigma_O) \times \frac{\Delta \Pi}{L}.    
\label{vxODOb}
\end{equation}

Now, adding the two contributions to the flux, $Q_{osm}$ and $Q_{DO}$, one gets the total flux as
\begin{align}
Q_w&=Q_{osm}+Q_{DO} \nonumber\\
&= \frac{k_{\rm hyd}}{\eta} \sigma_O {\Delta \Pi\over L} + {K}_{DO} (1-\sigma_O) \times \frac{\Delta \Pi}{L}.
\label{Qtot}
\end{align}
Defining a diffusio-osmotic reflection coefficient  as
$\sigma_{DO} = \frac{\eta {K}_{DO}}{k_{\rm hyd}}$ 
we then rewrite Eq.~\eqref{Qtot} as
\begin{equation}
 Q_w = \mathcal{L}_{{\rm hyd}} \left( \sigma_{DO} + \sigma_{O} - \sigma_{DO}\sigma_{O} \right) \Delta \Pi,
\label{additivity}
\end{equation}
where, as before, the permeance is defined  $\mathcal{L}_{\rm hyd} = k_{\rm hyd} /(\eta L)$.

The flow across the membrane still obeys a general Kedem--Kachalsky formula as in Eq.~\eqref{kk1}-\eqref{kk2}, with $\Pi$ the general osmotic force, but with a reflection coefficient
which now contains {the coupled effects of osmosis and diffusio-osmosis}:
%
\begin{equation}
 \sigma = \sigma_{O} + \sigma_{DO} -\sigma_{O}\sigma_{DO},
\label{sigmaTot}	
\end{equation}
hence defined in terms of the combination of  ``osmotic'' and ``diffusio-osmotic'' reflection coefficients. 
The expression for the global reflection coefficient has interesting limits to consider. For example, for 
completely semi-permeable with $\sigma_{O} = 1$,  the diffusio-osmotic contribution, which behaves as $\sigma_{DO} (1 - \sigma_O)$, vanishes. Similarly one also obtains $\sigma =1$ when $\sigma_{DO}=1$. 
The global effect of the combination of osmotic and diffusio-osmotic flow is however non-trivial in general, especially since $\sigma_{DO}$ can have a negative sign
-- which occurs when the solutes is depleted at the confining walls. 

This calculation relies on simplifying assumptions. But it  illustrates that osmotic and diffusio-osmotic transport will be in general intertwinned. For large nanochannels, no solute rejection is expected so that $\sigma_O\rightarrow 0$ and only the diffusio-osmotic transport remains. But for smaller nanochannels, where solute rejection and surface interfaction both apply, osmosis is {\it de facto} a mix from bare osmosis and diffusio-osmosis. 



\subsubsection{Combining various sources of rejection}

The above calculation is quite formal but
 it points to the possibility of combining the various mechanism leading to osmotic drivings, using the point of view of rejection coefficients. 
In general one may write
\beq
\Delta \Pi = \sigma k_BT \Delta c
\label{sigma-here}
\eeq
where the rejection coefficient $\sigma$ is expected to be a complex combination of all phenomena impeding solute transport: steric, electrostatic, surface adsorption, hydrodynamics, etc. These aspects have been discussed exhaustively in the literature, see \cite{anderson1974mechanism,hill1979osmosis} to cite a few, where the rejection is discussed in terms of the details of the solute-pore interactions. For example, a simple steric exclusion modelling of the solute of size $r_s$ in a pore of size $a$ would suggest that the steric rejection coefficient writes as ~\cite{anderson1974mechanism}:
\beq
\sigma_{\rm st}=\left[1-(1-\varepsilon)^2 \right]^2
\eeq
with $\varepsilon=r_s/a$ the ratio of solute to pore radii. 
Alernatively, in the presence of surface charge ($-e\Sigma$), there is  an electrostatic/Donnan contribution to the rejection coefficient, which we will find as, see Eq.(\ref{sigma-elec})
\beq
\sigma_{Donnan}(Du)=\left[1- \sqrt{1+Du^2}\right].
\label{sigma-elec0}
\eeq
with $Du=\Sigma/c_s h$, the Dukhin number with a pore of size $h$; see section VI.

While the various physical mechanism underlying the rejection coefficient $\sigma$ listed above would require a full microscopic description, one may propose a simple combination. First, it is interesting to remark that   
the effective rejection coefficient in Eq.(\ref{sigmaTot}) can be also understood in terms of the solute permeability coefficient $P$,
defined in terms of the solute flux across the nanochannel as $j_s= - D \times P\times \nabla c_{\rm sol}$. We anticipate from the discussion below that the permeability is intimately connected to the rejection coefficient $\sigma_0$ entering the osmotic expression -- cf Eqs.(\ref{Psigma0}) and (\ref{sigmaP}) --  via
\beq
P \propto 1-\sigma_0
\eeq
Under the simplifying assumption of the models described in Sec. III and below, the proportionality factor is one (although this may not be a general rule \cite{anderson1974mechanism}) and $P = 1-\sigma_0$; see Eq.(\ref{Psigma0}).
Now, since $P$ is related to the diffusion of the solute across the nanochannel, one may envisage $P$ as a probability for the solute to cross the channel. Then 
if various mechanisms affect the solute permeability independently, say phenomenon 1 ({\it e.g.} steric exclusion, ...) with rejection $\sigma_1$ and phenomenon 2 with rejection $\sigma_2$, it is expected that the overal permeability will be the product of the indivual permeabilities, similar to probabilities for independent variables. Here one expects accordingly
\beq
P_{\rm tot}=P_{1}\times P_{2}= (1-\sigma_{1})(1-\sigma_{2})
\label{Ptot}
\eeq
Now writing that $P_{\rm tot}=1-\sigma_{\rm tot}$, Eq.(\ref{Ptot}) thus implies that 
\beq
\sigma_{\rm tot}=\sigma_{1}+\sigma_{2}-\sigma_{1}\cdot\sigma_{2}.
\label{sigma_multiple}
\eeq
This result matches the one obtained above for the combination of osmotic and diffusio-osmotic flows in Eq.(\ref{sigmaTot}).

This should be considered merely a rule of thumb to combine the various underlying rejection mechanisms. But in general, a full  description including all the mechanism at play is required 
to account for the solute  and osmotic transport in the nanochannel. 


\subsection{Experimental observations of diffusio-osmosis}

Diffusio-osmosis and its consequences, in particular in nanochannels, have been reported in a number of experiments.
For example, ionic currents generated under diffusio-osmotic drivings have been measured in many different
configurations \cite{siria2013giant,pendse2021highly,pascual2023waste}, while in various studies 
consequences of DO transport were highlighted, such as  negative net osmotic flow \cite{lokesh2018}, 
DO effects inside an ion concentration polarization layer \cite{cho2016}, or its role in the emerging
 mechano-sensitive behavior in nanochannels with charge patterns \cite{Lizee2025architecting}.

\begin{figure}[h!]
    \centering
    \includegraphics[width=0.5\textwidth]{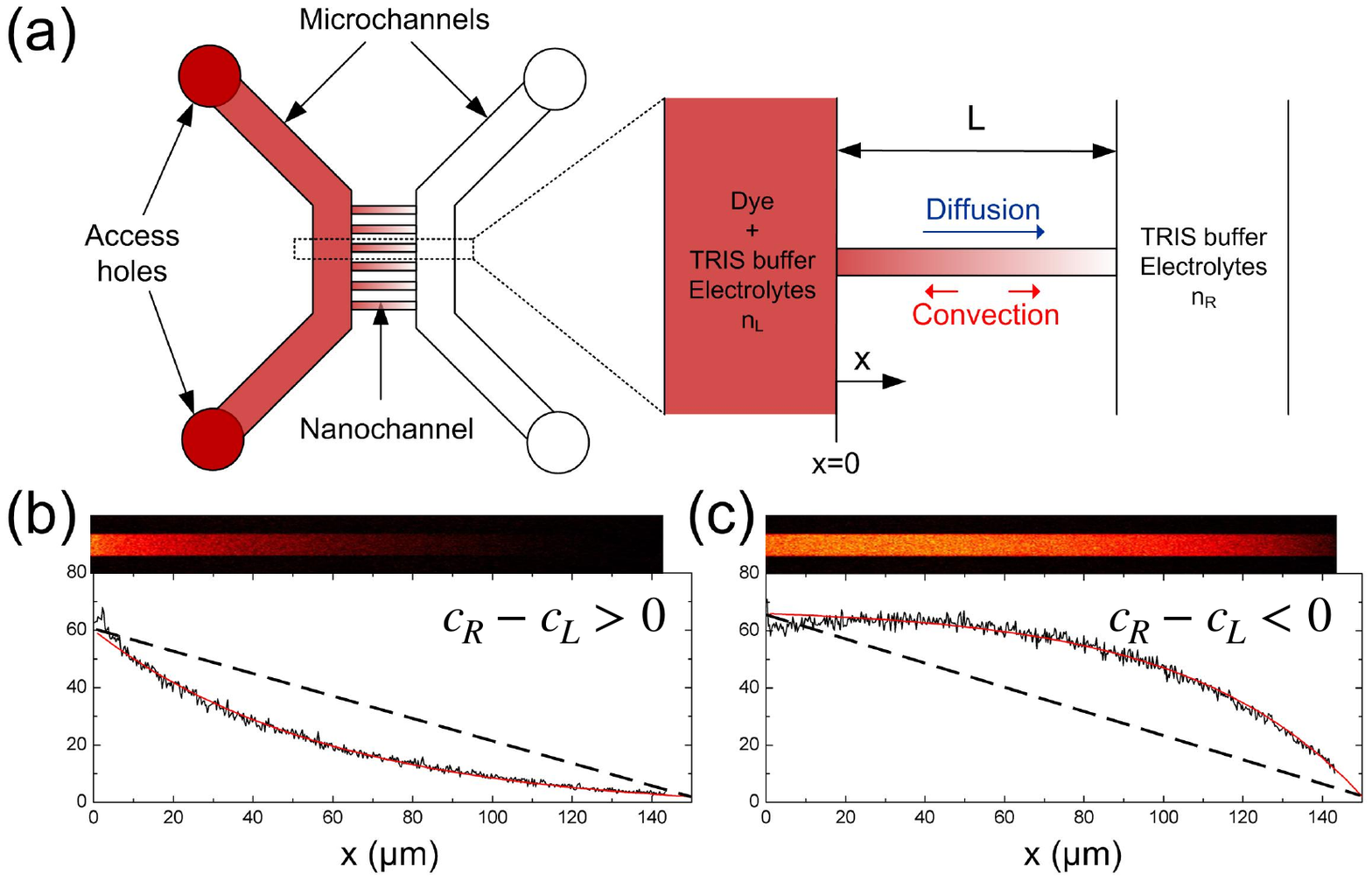}
    \caption{Experimental demonstration of DO flow, from Ref. \cite{lee2014osmotic}: 
        Nanochannels (height 163 nm, width  $w= 5 \mu$m, and length $L= 150 \mu$m) bridge
between two microchannels. 
in which solutions of different solute concentrations
fixing the solute gradient
along nanochannels. (b),(c) Steady state fluorescent probe
intensity profiles along nanochannels for different solute (NaI)
concentration gradient (b) The solute concentration contrast obeys $c_R-c_L >0$ (1 and 30 mM), while  in (c)  $c_R -c_L<0$ (30 and
1 mM). Solid line (red): fit according to diffusion-convection
equations (1); all profiles are seen to strongly depart from the
linear diffusive-only theoretical profile.
  Adapted from Ref. \cite{lee2014osmotic} with permission from American Physical Society (APS), Copyright (2014).}
    \label{fig:DOexperimentLee}
\end{figure}

However, there are very few {\it direct} measurements of DO, {\it i.e.} of a flow induced by a solute gradient. 
One such measurements is reported in \cite{lee2014osmotic,lee2017nanoscale},
in which DO transport along nanochannels is evidenced using fluorescence measurements. 
The experiments, whose results are highlighted in Fig.\ref{fig:DOexperimentLee}, studied
the flow in nanochannels submitted to a difference of concentration of a solute -- polymers, 
salts, mixtures -- at the two ends. The height of the nanochannel is $\sim 150$nm, so that the nanochannel is
{\it not} semi-permeable: there is no bare osmotic pressure under the concentration gradient. The flow induced by the concentration
gradient can only be explained by surface-driven diffusio-osmosis. 

Getting into the measurement, the DO flow is induced by a solute concentration difference at the two ends of the channel: $c_L$ at $x=0$ and $c_R$ at $x=L$. 
The flow probed by a fluorescent dye with minute concentration (so that it does not influence
DO velocity). The concentration profile of the  dye, say $c_f$, results
from the balance between dye diffusion and DO-induced convection, which writes
\begin{equation}
v_{DO}{\partial c_f\over \partial x} = D {\partial^2c_f\over \partial x^2} 
\end{equation}
(note that the surface-induced DO flow is merely plug-like across the section of the $150$nm channel).
Accordingly, $c_f$ thus take the characteristic shape
\beq
c_f(x)=c_L+(c_f^R-c_f^L)\times {e^{Pe_{DO}{x\over L}}-1\over e^{Pe_{DO}}-1}
\label{cDO}
\eeq
where $c_f^{R/L}$ are the dye concentration at the two ends of the channel (right and left) and the DO Peclet number is defined as
\begin{equation}
{\rm Pe}_{DO}={v_{DO} L \over D}
\end{equation}
which can be positive or negative depending on the direction of the DO velocity.

As shown on Fig.\ref{fig:DOexperimentLee}, experimental results for the dye concentration are perfectly reproduced by the prediction in Eq.(\ref{cDO}).
Furthermore the flow under a salt gradient is seen to be directed towards the lowest salt concentration, as expected
for DO flow, and in contrast to bare osmosis. These experiments have furthermore confirmed the linear
dependence of the DO velocity in $\Delta \log c_s$ for salts as a solute. In contrast, for a neutral solute (PEG polymer), the
DO velocity is proportional to $\Delta  c_{\rm sol}$, as $v_{DO}=K_{DO} k_BT (-\Delta c_{\rm sol})/L$, and  the flow is  towards
the highest concentration, hence reversed from the salt gradient case. This originates in the excluded volume of the polymer close to the surface, hence corresponding
to a repulsive situation for which the DO mobility is negative, with $K_{DO} =- R_g^2/\eta$ with $R_g$ the radius of gyration of the polymer.
For more complex mixtures, {\it e.g.} ethanol in water, the transport was found to be reversed as compared to what a naive 
adsorbing prediction would suggest \cite{lee2017nanoscale}. This originates in subtle dynamical effects of the solute close to the surface,
as discussed earlier. 

We shall discuss further experimental consequences of diffusio-osmotic transport in Sec. VIII. 


%


\section{A global force balance in osmotic transport across nanochannels }


In this section, we dig further into the mechanical couplings between entropic forces and mechanical drivings.  
We  derive a useful relationship, which connects the apparent osmotic pressure -- defined
as the ``push'' on the fluid under an (electro-) chemical gradient -- and the solute flux. This takes the form
of an integral expression, which we will use in various applications, see next sections. 


We consider a nanochannel (or an assembly of), 
assuming that the lateral dimensions of the nanopores are far smaller than their lengths, say $L$, 
and we accordingly simplifies the problem to that of a 1D geometry. 
A solute gradient is applied between the two sides of the nanochannel. 

Our goal is to show that an integral relation exists between the apparent osmotic pressure across the channel and
the solute flux $j_s$ (per unit surface) across the channel.
\beq
\Delta \Pi = k_BT \left[ \Delta c_{\rm sol} + j_s \times{ L \over D} \right]
\eeq
with $\Delta c_{\rm sol}$ the solute concentration difference.
This equation assumes a vanishing Peclet limit, $Pe=v_w\,L/D \ll 1$. A corrected expression taking into account  finite Peclet number effects 
will be also discussed. 

The derivation is quite similar in spirit for neutral and charged solutes.  
We will focus on neutral solutes to simplify the discussion, but the same results holds for ion transport, see Ref. \cite{picallo2013nanofluidic}.
The interaction of the solute with the pore, associated with its selectivity, is accounted for as a generic potential $\UU(x)$ along the channel. 

At a global scale,  the total water flux $Q$ is expected to be proportional to the pressure drop
\begin{equation}\label{fr}
Q=-{{\cal A}\over L}\,{k}_{\rm hyd} \times {\Delta[p-\Pi_{\rm app}]} ,
\end{equation}
with $k_{\rm hyd}=\mathcal{L}_{hyd}/{\cal A}$ the hydrodynamic permeability of the channel, and $\Delta\Pi_{app}$ the apparent osmotic pressure; 
 $\mathcal{A}$ is the channel cross section. Typically $k_{\rm hyd} \approx a^2/\eta$, with $a$ a length of order of the typical pore size (up to numerical prefactors) and $\eta$ the fluid viscosity. 
At a local scale inside the nanochannel, the fluid velocity responds linearly to the local forces acting on the fluid, as highlighted in the
 hydrodynamic (Stokes) equation, so that
\begin{equation}\label{eq_Fbal}
v_w={k}_{\rm hyd} ( -\nabla p + c_{\rm sol}\,(-\nabla \UU)),
\end{equation}
with 
$\UU$ the local  potential acting on the solute. Here we propose a 1D approach and neglect the lateral structure of the profiles across the nanochannel. For an electrolyte, the equivalent term would be the electric bulk force, $n_c (-\nabla V)$ with $n_c$ the charge density.

The last term in Eq.(\ref{eq_Fbal}) accounts for the force acting on the solute. Then,  following the spirit of the mechanical approach for osmosis described in Sec. III.A, one {\it defines} the apparent osmotic pressure in terms of this force as:
\begin{equation}\label{eq_Fc}
{\Delta\Pi_{\rm app}}=  \int_0^L dx\, c_{\rm sol}\,(-\nabla \UU) \equiv L\times \langle  c_{\rm sol}\,(-\nabla \UU)\rangle
\end{equation}
An average over the nanochannel cross section is also made implicitly.

Now, let us write the transport equation for the solute. We assume that it obeys the 
diffusion-convection equation \cite{bocquet2010nanofluidics} in a 1D geometry, with the various quantities averaged over the cross area. 
The solute flux writes as
\beq 
j_s = -D\nabla c_{\rm sol}+ \mu c_{\rm sol}\,  (-\nabla \UU) + \mu c_{\rm sol}\, v_w ,
\label{PNPS}
\eeq 
with $D$ the solute diffusion coefficient, ${\mu=D/ k_BT}$ the mobility and $v_w$ the averaged fluid velocity across the pore area.  
In a 1D geometry, mass conservation imposes that the flux $J_s$ is independent of $x$. The same applies for the mass flux, hence to the (averaged) velocity $v_w$.\\

{\it  Peclet number: diffusion versus advection --}
Let us quantify the effect of advection on the solute transport. In Eq.(\ref{PNPS}), the relative effects of convection compared to diffusion 
is quantified by a P\'eclet number
\beq
Pe=v_w\,L/D 
\eeq
Convection can be accordingly neglected for small P\'eclet number, $Pe=v_w\,L/D \ll 1$. 
This condition does not exclude the existence of a flow, but this flow does not influence the solute profile. 
Let's put numbers. The velocity under a pressure drop $\Delta P_{\rm tot}$ (including or not the osmotic pressure) is $v_w=-{a^2\over 8\eta} {\Delta p\over L}$ for a tube of radius $a$, length $L$; $\eta$ the viscosity of the fluid. I consider a nanochannel with radius below 10nm to fix ideas, say $a=2$nm, and length $L=10\mu$m. Accordingly the velocity is found in the $\mu$m.s$^{-1}$ range for a pressure drop $\Delta P_{\rm tot}=1$bar, and the Peclet number is $Pe\sim 5.10^{-2}$ hence small indeed. So nanofluidics is generally associated with small P\'eclet numbers. 
 
 We note however that large Peclet can still arise in the presence of hydrodynamic slippage at the walls of the confining nanochannels since the permeability -- hence the flow -- is strongly increased  by slippage. Such effects occur for example in carbon nanotubes or graphitic nanochannels \cite{secchi2016massive,mouterde2019molecular}. 
We come back below on this hypothesis, which can have important  consequences on the transport, in particular the emergence of non-linear, mechano-sensitive effects \cite{secchi2016massive,marcotte2020mechanically,mouterde2019molecular}.\\


{\it Small Peclet regime: neglecting advection --}
Let us first consider the small P\'eclet limit and neglect the  convective contribution in Eq.(\ref{PNPS}).
In the stationary state, the 1D flux is homogeneous in space and time, and $j_s = -D\nabla c_{\rm sol}+ \mu c_{\rm sol}  (-\nabla \UU) = {\rm cst}$. 
This allows rewriting the total  force acting on the solute  in terms of $j_s$ and $\nabla c_{\rm sol}$, as 
\beq
\int_0^L dx\, c_{\rm sol}\,(-\nabla \UU)={L \over \mu} \, j_s +{D\over \mu}  \int_0^L dx\, \partial_x c_{\rm sol}, 
\eeq
so that 
the apparent osmotic pressure in Eq. (\ref{eq_Fc}) now writes
\begin{equation}\label{eq_Piapp}
\Delta\Pi_{\rm app}=k_BT\left(\Delta c_{\rm sol} +{j_s}\times{L\over D}  \right) , 
\end{equation}
where $\Delta c_{\rm sol}$ is the solute concentration difference. 

This integrated expression makes an explicit connection between the osmotic pressure and the solute flux across the nanochannel. It has some interesting limiting regimes.
For a semi-permeable membrane, the solute flux vanishes, $j_s = 0$, and the previous equation reduces to $\Delta\Pi_{app}=k_BT\, \Delta c_{\rm sol}$, {\it i.e.} matches the 
van 't Hoff expression. 
For a fully permeable  channel, 
{$j_s=-D\,\Delta c_{\rm sol}/L$} and the osmotic vanishes: $\Delta\Pi_{app}=0$, as  expected.
In general the nanochannel is only partly permselective and $\Delta\Pi_{app}$ takes a non-vanishing value \emph{depending on the solute flux}. 
%
If one writes $j_s= - P \times D \Delta c_{\rm sol}/L$ with $P$ a solute permeability coefficient across the porous material, then
\beq
\Delta \Pi = \sigma k_BT \Delta c_{\rm sol}
\label{sigma}
\eeq
with the reflection coefficient 
\beq
\sigma = 1-P
\label{sigmaP}
\eeq 
as previously obtained.

{\it Regime of small but non-vanishing Peclet --}

This result extends to finite P\'eclet, although at the expense of some further approximation, and therefore a somewhat reduced generality. 
From the hydrodynamic equation, Eq.(\ref{eq_Fbal}), one may still write 
\begin{equation}\label{fr2}
v_w={k}_{\rm hyd} \times {\left[-{\Delta p\over L}+\langle c_{\rm sol}\,(-\nabla \UU)\rangle\right]} 
\end{equation}
and the averaged ``membrane force'' $\langle c\,(-\nabla \UU)\rangle$ is now  
\begin{eqnarray}
&\langle c_{\rm sol}\,(-\nabla \UU)\rangle&={1\over L} \int_0^L dx\,  c_{\rm sol} (-\nabla \UU) \label{piapp_conv} \\
&&={1\over L} \times k_BT\left( \Delta c_{\rm sol} +  {L\over D}\times j_s  -{L  \over D} \langle c_{\rm sol}\, v_w \rangle  \right)\nonumber
\end{eqnarray}
where the second line follows from the diffusion-convection equation in Eq. (\ref{PNPS}); $v_w=Q/{\cal A}$.

Now, at finite Peclet, the right-hand side of Eq.(\ref{piapp_conv}) involves convective terms proportional to the fluid velocity $v_w$.
In general, one should calculate explicitly the solute profile $c_{\rm sol}(x)$ in order to estimate this contribution. However
these terms essentially renormalize the fluid permeability. Indeed, following \cite{manning1968binary}, one may write generally $j_s = -D\times P{ \Delta c_{\rm sol} \over L}+ \alpha\, v_w$, 
with $P$ the diffusive permeability and $\alpha$ a transport coefficient gathering all convective contribtions to $j_s$ (see for example the result for $j_s$ in Eq.(\ref{fluxfinal}) in the thick diffuse layer regime). This
 term $ \alpha\, v_w$ renormalizes the global flux, and modifies accordingly the permeabilty of the channel. 
Eq.(\ref{fr2}) is then rewritten as
\beq
v_w=-{{\tilde{k}}_{\rm hyd} \over L} \left[\Delta p -k_BT(1-P)\Delta c_{\rm sol}\right]
\eeq
with $\tilde{k}_{\rm hyd}={{k}_{\rm hyd}\over 1 +{k}_{\rm hyd}  {k_BT\over D}(\langle c_{\rm sol} \rangle-\alpha)}$ a renormalized permeability; $P$ is the solute
permeability.  \\

{\it Apparent osmotic pressure --}
Altogether we conclude that the apparent osmotic pressure takes the expression 
\beq
\Delta\Pi_{\rm app}=k_BT\left(\Delta c_{\rm sol} +j_s^{\rm n.c.} \times{L\over D}  \right) , 
\label{DPi}
\eeq
with $j_s^{\rm n.c.}={j_s}- c_{in} v_w$  is the non-convective part of the solute flux; $c_{in}=\langle c_{\rm sol}\rangle$ the average solute concentration inside the pore. 
The terminology of the ``apparent'' osmotic pressure points to the fact that $\Delta \Pi_{\rm app}$ will be the quantity measured in an experiment where one measures the fluid flow under a solute/salinity gradient, according to Eq.(\ref{fr}). The ``apparent'' osmotic pressure is in general a combination of various contributions, osmotic, diffusio- and electro- osmotic, etc. which are difficult to disentangle. In the following sections, we will explicitly discuss these mixed contributions to osmosis. 
However what is measured experimentally is their overall (mechanical) contribution to the fluid flow. Hence, as a ``push'' on the fluid under a solute gradient, it desserves the terminology of an osmotic force.

We will illustrate several examples of the use of the integral relation in Eq. (\ref{DPi}) in the sections below. It is particularly useful to obtain the osmotic pressure in complex geometries, as well in situations where non-linear osmotic transport occurs.
Indeed one may note that the integral relationship does not assume linear transport. Hence it can be used to explore non-linear osmotic and diffusio-/electro-osmotic effects, as we will discuss in the context of the osmotic diode in Sec. VIII.

\section{Local electrokinetic and osmotic transport in nanochannels: the capillary pore model}

Let us focus on salts as a solute. 
Osmotic drivings -- be it bare osmosis or DO -- are part of a larger electrokinetic scheme, where
transport occurs under the coupled drivings of mechanical (pressure) forces, electric forces
and electro-chemical gradients. The profiles of the salt/solute concentration, pressure and electric potential
are intimately coupled, often in a very non-linear way. 

 The global transport output, as summarized in the matrix ${\mathbb L}$ in Eq.(\ref{Larray2}), thus results
from the combination of all these transport mechanisms. 
These couplings offer a considerable wealth of behaviors, resulting in
both linear and non-linear transport, hence making transport inside nanochannel  inherently complex and rich.
The case studies  in Sec. VIII will examplify a few examples.

\subsection{PNP+Stokes framework}

%

The usual starting point to describe the local transport in the nanochannel (or a collection of) is the Poisson-Nernst-Planck formalism, coupled to the Stokes equation, which, in the stationary state, writes as
\begin{eqnarray}
&\nabla\cdot j_{\pm}=0 \nonumber \\
&j_{\pm} = -D_\pm\left( \nabla c_{\pm} \pm \frac{e}{k_B T} c_{\pm} {\nabla V} \right)+v_w c_\pm
\label{PNP-1}
\end{eqnarray}
\begin{equation}
0=-\nabla p +\eta \Delta v_w -e(c_+-c_-) {\nabla V} 
\label{Stokes2}
\end{equation}
where for simplicity I assumed monovalent ions and $D_\pm=D$ (both hypothesis are easily relaxed).
The time scale for the relaxation within the diffuse layer is in the range of $\lambda_D^2/D \sim$ns, so that the electric double layer can be usually assumed to relax to its (local) equilibrium. The transport equations thus reduce to the relaxation of the fields {\it along} the channels' length. 
This description constitutes the core of the so-called capillary pore model, or Space-Charge Model, 
introduced by Osterle and his collaborators \cite{gross1968membrane,fair1971reverse,westermann1983experimental} and revisited recently in Ref.  \cite{peters2016analysis}.  
I recall here this description since it proves usefull to describe nanoscale transport, and in particular osmotic drivings. 

The description reduces to the linear relationship for the {\it local} fluxes, averaged over the cross-section of the channel,  versus the {\it local} forces along the channel length, say $x$, in the form
\beq
\begin{pmatrix}
{\langle v_w\rangle} \\
{\langle j_{\mathrm{s}}\rangle}-c_v \langle v_w\rangle \\
{\langle i_{\mathrm{e}}\rangle}
\end{pmatrix}
= 
\begin{pmatrix}
L_{11} & L_{12} & L_{13} \\
L_{21} & L_{22} & L_{23} \\
L_{31} & L_{32} & L_{33}
\end{pmatrix}
\begin{pmatrix}
-\nabla_x [p_{v}-\Pi[c_v]] \\
-\nabla_x  \mu_v\\
-\nabla_x \phi_v
\end{pmatrix}
\label{Lmatrix}
\eeq
where $p_v$ is the local hydrodynamic pressure, $\Pi[c_v]=k_BT c_v$ the local osmotic pressure, $\mu_v=k_BT \log[c_v v_0]$ the chemical potential, $\phi_v$ the axial potential along the channel length (defined below), and the fluxes are averaged over the cross-section of the channel.  The index ``$v$'' stands for ``virtual'', as I discuss below. Note that the introduction of the osmotic pressure in the driving force $-\nabla_x [p_{v}-\Pi[c_v]]$ is merely a convention at this stage. For example, in the thin diffuse layer regime, the bare osmotic contribution is expected to cancel out in the final result for the water flux (keeping only the surface, diffusio-osmotic, contribution).

This framework introduces the local matrix of transport, $\mathbf{[L]}$, where each term $L_{ij}$ should be calculated on the basis of the PNP+Stokes equations. 
Complete expressions for the $L_{ij}$ can be found in \cite{fair1971reverse,peters2016analysis}
in terms of integrals of the PB electrostatic potential. For illustration in this chapter, I will only mention the expression 
for the osmotic cross term, $L_{12}=L_{21}$ which, in a circular pore geometry, takes the expression
\beq
L_{21} = L_{12} = {k_BT\over \eta } c_v \int_0^R dr\, r\left(1 - \left({r\over R}\right)^2\right) \cosh \psi(r)
\eeq
 where $\psi(r)$ is the dimensionless PB electrostatic potential. Expressions for the other terms can be quite cumbersome.
 
In the following sections, I will discuss in details the two limiting regimes of thick and thin Debye layers, where explicit expression can be obtained for all terms of the $\mathbf{[L]}$ transport matrix. 

But let me first dig further into the capillary pore formalism. 


%


\subsection{Main steps of the derivation}
The derivation of these results is interesting {\it per se} and let me give the main lines of the derivation
of such results. This followis the presentation in Ref.\cite{peters2016analysis}. I will use the cylindrical pore geometry
as in \cite{peters2016analysis}, but the analysis can be easily extended to any other. 

The underlying idea is to separate the variations along the channel length ($x$) and in the radial direction ($r$). 
Similarly to Eq.(\ref{Usep}), this is done by separating the electric potential as
\beq
 V(x, r)= \phi_v(x) + \tilde\psi(x, r)
\label{Sepxy}
\eeq
  where, as explained in Refs. \cite{fair1971reverse,peters2016analysis}, the radial potential $\tilde\psi(x,r)$ is obtained from an
equilibrium PB model, and $\phi_v(x)$ accounts for axial gradients in potential along the length of the pore.
For convenience I introduce the dimensionless radial potential $\psi={e\tilde\psi\over k_BT}$.

A key difficulty is that in full generality, the potential in the middle of the nanochannel does not vanish, and there
is partial or total overlap of the electric double layers. Accordingly there is no ``bulk region'' in the channel far from
the boundaries, in particular for the concentration. To circumvent this difficulty, the trick is to introduce a virtual reservoir 
which is in equilibrium with any slice in the pore \cite{peters2016analysis}. In particular the lateral Boltzmann distribution for the
ion concentrations verifies $c_\pm(x,r)=c_v(x) \exp[\mp \psi(x,r)]$ and the Poisson-Boltzmann equation
writes 
\beq
\Delta \psi = {8 \pi \ell_B c_v(x)} \sinh \psi(x, r).
\eeq
where $\Delta$ is the Laplacian, which in the cylindrical nanopore takes the expression $\Delta \psi \equiv \frac{1}{r} \frac{\partial}{\partial r} \left( r \frac{\partial \psi(x, r)}{\partial r} \right)
$. The local Debye length is defined as $\lambda_D[c_v(x)]=({8 \pi \ell_B c_v(x)})^{-1/2}$. 
At equilibrium, {\it i.e.} when all driving forces are zero, the virtual concentration $c_v$ identifies with that  
of the reservoir since in PB $c_v={v}_0^{-1}\exp[\beta \mu_{\rm res}]\equiv c_{\rm res}$ (${v}_0$ an irrelevant
molecular volume).
When driven out-of-equilibrium, for example when the concentrations in the two reservoirs differ, a profile 
of virtual concentration builds up. 
Similar considerations apply for the pressure profile. Due to the conditions of local equilibrium in the cross section of the channel, 
the hydrostatic pressure can be written as 
\beq 
p(x,r)=p_v(x) + 2k_BT c_v(x) (\cosh \psi -1), 
\label{hydropv}
\eeq
where the term
$p_v(x)$ only depends on $x$ and is coined the virtual pressure. This is exactly the same conclusion as for Eq.(\ref{hydrostat}).

The salt flux $j_s(x,t)= j_+(x,r)+j_-(x,r)$ and current density $i_e= j_+(x,r)-j_-(x,r)$ along $x$ are found to reduce to simple expressions in terms
of the perpendicular electrostatic field \cite{peters2016analysis}:
\begin{eqnarray}
j_{s}(x,r) &= 2c_v \cosh\psi(r)\,u_x(r)
- {D\over k_BT} 2c_v \cosh\psi(r)\,\frac{\partial \mu_v}{\partial x}(x)\nonumber \\
&+ {eD\over k_BT} 2c_v \sinh\psi(r)\,\frac{\partial \phi_v}{\partial x}(x), \\
i_{e}(x,r) &= - 2c_v \sinh\psi(r)\,u_x(r)
+ {D\over k_BT} 2c_v \sinh\psi(r)\,\frac{\partial \mu_v}{\partial x}(x)\nonumber \\
&- {eD\over k_BT}2c_v \cosh\psi(r)\,\frac{\partial \phi_v}{\partial x}(x).
\label{Capmodel-equ}
\end{eqnarray}
On the other hand, the velocity field $v_x(r)$ is obtained in terms of the force fields $-\nabla_x [p_{v}-\Pi[c_v]]$, $-\nabla_x  \mu_v$,
$-\nabla_x \phi_v$ by integrating the Stokes equation across the diffuse layer, in a very similar way as developped in Sec. III.
Here the Stokes equation is rewritten is a compact form as \cite{peters2016analysis}:
\begin{eqnarray}
&\eta \, \Delta_r v_x(r)
&= -\frac{\partial [p_{v}-\Pi[c_v]]}{\partial x} 
+ 2 c_v \cosh \psi(r) \frac{\partial \mu_v}{\partial x}\nonumber\\
&&- 2 e c_v \sinh \psi(r) \frac{\partial \phi_v}{\partial x} .
\label{Capillary}
\end{eqnarray}
with $\Delta_r=\frac{1}{r} \frac{\partial}{\partial r}
\left( r \frac{\partial }{\partial r} \right)$ the cylindrical Laplace operator. 

Once integrated, the averaged flux, current and flow are calculated in terms of the driving forces and one  deduces  the expressions for the various terms $L_{ij}$ of the local transport matrix.

\subsection{Global transport}

Going from the local transport matrix in Eq.(\ref{Lmatrix}) to the global transport is by no mean an obvious journey. 
The local transport equations in Eq.(\ref{Lmatrix}) should be solved together with proper boundary conditions at the 
nanochannels ends. Typically in the capillary pore model, one assumes that the potential at the nanochannels ends is that the virtual potential $\phi_v$ matches the potential in the reservoirs and the virtual pressure matches the hydrostatic pressure in the reservoirs. Note however that alternative boundary conditions apply for ``real'' fields different from the ``virtual'' ones defined there: we refer in particular to the description of the thick layer regime introducing the Donnan potential in Sec. VI. 

The results of the model are exhaustive, since they embrace {\it de facto} all electro-kinetic transport
phenomena: electro-osmosis, diffusio-osmosis, etc. But they are {\it per se} difficult to exploit. 
First the expressions for the transport coefficient are quite opaque and can only be calculated numerically
in the last steps for some of the terms $L_{ij}$. However analytical expressions for the $L_{ij}$'s can be obtained in the thin and thick layers regimes, see Sec. VI and VII.

When linearized, the resolution of the local transport equation leads to a linear global relationship, as introduced in the transport matrix introduced in Eq.(\ref{Larray2}); see for example a solved example for the thick diffuse layer regime in Eq. (\ref{Tmatrix-PNP}). 
However, locally linear transport does not imply global linear transport and non-linear global transport can emerge; see {\it e.g.} Sec. VIII.C\&D. 
Non-linearities result usually from the couplings of the various electrolyte transport mechanisms: flushing, charge and concentration polarization, etc. For example, inhomogeneous surface
charge patterns on the surface of the nanochannel are know to yield ionic diodes of the Schokley type \cite{karnik2007rectification}, and even ``osmotic diodes'' as predicted in \cite{picallo2013nanofluidic} and measured
recently in composite membranes \cite{abdelghani2025resonant}. Similarly mechano-sensitive effects
in the form of pressure dependent transport properties were measured in nanochannels \cite{marcotte2020mechanically,mouterde2019molecular}.
The latter are usually observed in the presence of  hydrodynamic slippage on the nanocapillary surface,
leading to high Peclet number (strong flush) and associated consequences.

As a final comment, we mention that the capillary pore model has been thoroughly compared to experimental data in the work by Westermann-Clark and
Anderson \cite{westermann1983experimental}. The authors used track-etched mica membranes 
whose pores were uniform capillaries in the range 34-265$\AA$ radius (and long pore lengths greater 
than 7 $\mu m$). They compared successfully their measurement of the various electrokinetic transport  
to the prediction of the capillary pore model, hence validating its foundations.

%
%
%
%
%


\section{Overlapping Double layers : mixed osmosis across nanochannels}

The regime of thick diffuse layers corresponds to the situation where the diffuse layers over each surface overlap,  
{\it i.e.} the pore size is smaller than (twice) the Debye length.
This limit  is   experimentally relevant for small nanochannels (and/or small salt concentrations) and  it provides a relevant guide into the 
understanding of complex experimental results. It is also an 
interesting regime where one can perform calculations explicitly, therefore allowing us 
to dig into the subtle couplings between the various EK phenomena, and in particular
the emergence of entropic forces.

In particular, it allows calculating analytically the global fluxes -- as summarized in the global transport matrix in Eq.(\ref{Larray2}) -- 
starting from the local relationships -- as introduced for example in the capillary pore model in Eq.(\ref{Lmatrix}).  


\subsection{The simplified PNP transport framework (without flow)}

We first discuss a simplified (but useful) approach of transport in nanochannel illustrating 
how the coupled transport emerge \cite{bocquet2010nanofluidics}. This is  the opportunity to introduce the Donnan potential, a key quantity for ion equilibrium and transport across nanochannels. The calculation illustrates the basic phenomena at the origin of the coupling between the various electrokinetic transport and I will thus enter the calculations with some details in the next paragraphs. 

Also, as a first discussion and for the sake of simplicity, I will focus here on 
\{electric + chemical \} drivings, in a strongly confined 1D nanochannel, omitting for now the mechanical drivings (and convection), which we will discuss in the next section. 
The channel has a slit shape with height $h$, width $w$ and length $L$. In the strongly confined nanochannel, we will further neglect the variations  of the ionic profiles in the transverse direction (perpendicular to the channel length) and thus reduce a 1D geometry.

{\it Equilibrium --}
In a charged nanochannel, an electric potential profile builds up at equilibrium due to the condition
of electroneutrality, which imposes that the positive and negative ion concentration in the nanochannel
should differ by the surface charge contribution: 
\beq
c_+-c_-={2\Sigma\over h}
\eeq 
for a negative surface charge density $-e\Sigma$ on the walls. Accordingly, a potential 
builds up inside the channel at equilibrium, obeying that  the ion chemical potentials equal that of the reservoir: 
\beq 
\mu_{\pm}=k_BT\log c_\pm \pm e V_D = k_BT \log c_s,
\eeq 
with $V_D$ the Donnan potential 
and $c_s$ the salt concentration in the reservoir ($c_s=c_+^{\rm res}=c_-^{\rm res}$). The ion concentrations inside the channel thus
obey the two equations: $c_+\times c_-=c_s^2$, and $c_+-c_-={2\Sigma\over h}$.
%
Hence
\beq
c_{\pm} [c_s]= c_s\left[ \sqrt{1 + Du^2} \pm Du\right]
\label{cboundary}
\eeq

\beq
V_D[c_s] =- \frac{k_B T}{e} \log \left[ {\sqrt{1 + Du^2} + Du} \right]
\label{Vboundary}
\eeq
with the Dukhin number defined as
\beq
Du= {\Sigma \over c_s h}
\eeq

{\it Transport equations --}
Let us now impose a different  salt concentration $\Delta c_s$ and electric potential $\Delta V$ in the reservoirs between the two ends of the nanochannel.
In line with the equilibrium calculation above,  there is a Donnan discontinuity of the electrostatic potential and salt concentration when entering the pore from both sides, due to the surface charge $\Sigma$ on the pore walls. 
Salt concentration and potential drops occur at the entrance and exit of  nanochannel, Eqs.(\ref{cboundary})-(\ref{Vboundary}) for the corresponding salt concentration  and voltage at the boundaries.

In 1D, the transport equations reduce to
\[
\partial_x j_\pm = 0
\]
where the ion fluxes take the expressions
\begin{equation}
j_\pm =-D_\pm \left(\partial_x c_\pm \pm  {e\over k_BT} c_\pm (-\partial_x V) \right)
\end{equation}
This is complemented by the assumption of  local electroneutrality, $ c_+-c_-=2\Sigma/h$.  

Let's introduce the ion flux and current densities (per unit surface), $j_s=J_s/{\cal A}=(j_++j_-)$ and $i_e=I_e/{e\cal A}=(j_+-j_-)$. One obtains 
\begin{equation}
j_s=-D\partial_x c + {D\over k_BT} {2\Sigma\over h} e\, (-\partial_x V) 
\label{flux}
\end{equation}
\begin{equation}
i_e= {D\over k_BT}  c \, e(-\partial_x V)
\label{current}
\end{equation}
where $c=c_++c_-$ is the total (local) ion concentration; for simplicity we assumed $D_+=D_-=D$. 

Conservation of ions imposes that both the flux $j_s$ and the current $i_e$ are spatially homogeneous over the channel length. To first order in $\Delta c_s$ and $\Delta V$, the concentration and electric potential profiles are linear inside the channel, so that $\partial_x c \simeq [\Delta c]_{\mathrm{in}} /L$ and $\partial_x V= [\Delta V]_{\mathrm{in}}/L$, defined in terms of the field difference  between the two ends {\it inside} the nanochannel. A simple calculation yields
\begin{equation}
[\Delta c]_{\mathrm{in}} = \frac{1}{\sqrt{1+Du^2}} \, 2 \Delta c_s,
\end{equation}
and
\begin{equation}
[\Delta V]_{\mathrm{in}} = \Delta V + \frac{k_B T}{e} \frac{{Du}}{ \sqrt{1+Du^2}} \, {\Delta c_s\over c_s}.
\end{equation}

Inserting in the flux equations, Eqs.(\ref{flux})-(\ref{current}), one gets finally global transport matrix as
\beq
\begin{bmatrix}
I_e\\
J_s
\end{bmatrix}
=\;
\;
{\mathcal{A}\over L}
\begin{bmatrix}
K & \mu_K \\
\mu_K & \mu_{\mathrm{eff}}
\end{bmatrix}
\times 
\begin{bmatrix}
{-\Delta V} \\
- k_B T \, {\Delta \left[ \log c_s \right]}
\end{bmatrix}
\label{Tmatrix-PNP}
\eeq
with $\cal A$ the cross section of the channel ($I_e$ and $J_s$ the integrated current and ion flux). 
The expressions for the coefficients of the  transport matrix are given as 
\begin{align}
&K = 2 \mu e^2 c_s \sqrt{1+Du^2},\nonumber \\
&\mu_{\mathrm{eff}} = 2\mu c_s\, \sqrt{1+Du^2}, \nonumber\\ 
&\mu_K = 2 e \mu \,  c_s \times Du 
\label{PNP-Donnan}
\end{align}

Note that due to Onsager (time-reversal) symmetry, the non-diagonal coefficients of the matrix are equal.
This result highlights how the Donnan potential modifies the transport inside the nanochannel,
with $\Du$ dependent transport accounting for the surface charge. 
Some interesting outcomes already emerge: for example, it predicts that the diffusion permeability is $P=\sqrt{ 1 + \Du^2 }$ -- with a counter-ion increased transport and a co-ion hindered transport --; 
it also predicts that an electric current is induced under the chemical gradient, 
with 
\beq
I_e^{\rm osm}={{\cal A}\over L} eD  {2\Sigma\over h} \Delta \left[ \log c_s \right]
\label{Iosm-thick}
\eeq
This is the signature of an osmotic ion current and we will come back extensively on this effect below, see Sec. VIII.D.

%
%
%

\subsection{The capillary pore with uniform potential reduces to the PNP-S model}

\subsubsection{Transport and boundary conditions}
In the limit of overlapping EDL, the equations for the capillary pore reduce to simple and transparent expressions which identify
with the Poisson-Nernst-Planck-Stokes (PNP-S) equations in Eq.(\ref{PNPS}) supplemented by a local electroneutrality assumption \cite{peters2016analysis}:
\begin{eqnarray}
&v_w =& {R^2\over 8\eta} \left[-{\partial_x p^h} +{ 2e\Sigma \over R} {(-\partial_x V)} \right], \label{eq1} \\[10pt]
&j_{{s}} =& - D {\partial_x c} + {D\over k_BT} e{ 2\Sigma \over R}(-{\partial_x V})+c\times  v_w ,  \label{eq2}\\[10pt]
&i_{{e}} = & e{D\over k_BT}\, c \times (-{\partial_x V}) +  {2\Sigma \over R} \times v_w \label{eq3}, 
\end{eqnarray}
where $v_w$ is the averaged flow velocity along the channel, $j_{\text{s}}=J_s/{\cal A}$ the total ion flux (per unit surface) and $i_{\text{e}}=I_e/(e{\cal A})$ the ionic current density (per unit surface); $\sigma =-e\Sigma$ is the surface charge on the channel. For simplicity I assumed $\Sigma$ to be homogeneous and discarded a $\partial \Sigma/\partial x$ term.
In these equations, $c=c_++c_-$ is the total ion concentration. 
In line with the capillary pore model discussed above, the geometry is assumed here to be that of a cylindrical nanopore. But the calculations can be easily extended to any pore geometry. 

A no-slip boundary condition is assumed here, a condition that can be relaxed by introducing a finite slip length on the channel surface, enhancing the permeability by a prefactor $1+{4v\over R}$ (of a cylindrical geometry). Such finite slippage effect have been observed in carbon nanotubes \cite{secchi2016massive,li2025carbon} and 2D slits and can lead to strong flushing effect under flow with  important consequences on the overall electrokinetic response
\cite{marcotte2020mechanically,mouterde2019molecular,Lizee2025architecting}.

As a side note, Eq.(\ref{eq1}) does {\it not} contain any explicit contribution from the osmotic pressure, {\it i.e.} a term proportionnal to $\partial_x \Pi[c]$, with $\Pi[c]=k_BT c$. This absence is {\it a priori} expected from a direct analysis of the Stokes equation for the water transport. But this is in fact by no means obvious when starting from Eqs.(\ref{Capillary}). It can however be  verified that this (osmotic) term indeed cancels out from the velocity equation in the thick EDL limit \cite{peters2016analysis}, and the flow velocity reduces to Eq.(\ref{eq1}). However, as we discuss extensively below, in spite of this absence in the local transport equation (\ref{eq1}), there is an overall osmotic pressure which builds across the channel, associated with the behavior at the boundaries of the channel. 

A further remark is that ions can interact with the confinement via other mechanisms than the electrostatic interactions under scrutiny here (say, dielectric, steric, etc., see section III.D.2). Such supplementary mechanisms may contribute to the transport equation via local interactions, leading to specific osmotic terms, so that an osmotic contribution may {\it in fine} appear in Eq.(\ref{eq1}, associated with the corresponding rejection mechanism.


At the boundary of the channels, one has to write the Donnan equilibrium relating the various parameters (concentration, etc.) inside the channel
to their values in the reservoirs. As discussed in the previous section, the Donnan conditions write for the  total ion concentration $c=c_++c_-$ and potential $V$ as 
\begin{eqnarray}
&c_{in} = \sqrt{ \left({ 2\Sigma \over R}\right)^2 + c_{\text{ext}}^2}=2c_s\sqrt{1+Du^2}\, , \label{eqDonnan2}\\[10pt]
&V_{in} - V_{\text{ext}} =- \frac{k_B T}{e} \log \left[ {\sqrt{1 + Du^2} + Du} \right]\label{eqDonnan3}
\end{eqnarray}
with $c_s$ the salt concentration. We further add a condition for the hydrostatic pressure $p^h$,
\beq
p_{in}^h = p_{\text{ext}}^h + k_BT \left(c_{in} - c_{\text{ext}}\right), \label{eqDonnan1}
\eeq
Here the subscript ``in'' refers to a position just within the membrane, 
and  ``ext''  a position just outside the membrane, in the (left or right) reservoir; $c_{\text{ext}}= [c_++c_-]_{\text{ext}}=2c_s$ is the total ion concentration in the reservoirs.
These conditions account for the discontinuity in the various fields (in reality a quick transition) which occurs at the nanochannel entrance  due to the presence of the surface charge inside the nanochannel.
These equations naturally introduce the Dukhin number \cite{bocquet2010nanofluidics}, which we define here as
\beq
Du= {\Sigma \over R\, c_s}
\eeq

It is interesting to dig somewhat into the condition for the hydrostatic pressure in Eq.(\ref{eqDonnan1}). Extending on the idea of a local equilibrium, this condition stems from the force balance at the entrance: the difference of electric potential at the interface associated with the Donnan potential should be compensated by a pressure and/or salt concentration difference.  
 To obtain it, one may integrate the 
Navier-Stokes equation along $x$ across a discontinuity at the channel's entrance (say at $x=0$) to obtain that 
\beq
0=\int_{0^-}^{0^+} dx \left[ -\partial_x p^h + n_c (-\partial_x V) \right]
\eeq
where $n_c=e(n_+-n_-)$  is the local charge density. The other terms in the equation vanishes because of equilibrium. 
Similarly integrating the total ion flux, which is also vanishing at equilibrium, leads to 
\beq
0=\int_{0^-}^{0^+} dx \left[  -D\partial_x c + {D\over k_BT} \times  n_c (-\partial_x V) \right]
\eeq
The second equation shows that the
electric driving associated with the interface electric field is compensated by the concentration difference:
$\int_{0^-}^{0^+} dx \left[  {D\over k_BT} \times  n_c (-\partial_x V) \right]=k_BT (c_{in}-c_{ext})$. 
This result  directly echoes  the global force balance discussed in Sec. IV. Gathering the results, 
we thus deduce the boundary condition for the pressure difference as
$p_{in}^h-p_{ext}^h=k_BT (c_{in}-c_{ext})$, as written in Eq.(\ref{eqDonnan1}).

\subsubsection{Apparent osmotic pressure}

Let us focus on a situation where a salt concentration difference is applied across the channel, {\it i.e.}  $c_{\text{ext}}^R=c_s+\Delta c_s $, $c_{\text{ext}}^L=c_s$ ($R$ and $L$ stands for the right and left reservoirs).
The concentration gradient in the reservoirs yields a concentration difference inside the channel, but also an electric potential difference and a hydrostatic pressure difference, according to Eqs.(\ref{eqDonnan2}-\ref{eqDonnan1}).

Furthermore, for small driving forces, the inner profiles for the concentration, pressure and electric potential will be essentially linear functions of position. Accordingly,   
in the capillary pore or PNP-S equations in Eqs.(\ref{eq1}-\ref{eq3}), the fluxes and corresponding gradients are defined in terms of the quantities calculated at the extremeties, but {\it inside} the channel: 
\begin{eqnarray}
&\partial_x c &= [c_{in}^R-c_{in}^L]/L, \nonumber \\
&\partial_x p^h &= k_BT \left[(c_{in}^R-c_{\rm ext}^R)-(c_{in}^L-c_{\rm ext}^R)\right]/L, \nonumber \\
&\partial_x V& = [V_{in}^R-V_{in}^L]/L, 
\end{eqnarray}
with $L$ the length of the channel; $c_{in}^{R/L}$ corresponds to the total ion concentration in the reservoirs. Remind that these are twice the corresponding salt concentration $c_s^{R/L}$.

As a consequence of Eq.(\ref{eq1}), the difference of concentration in the reservoirs leads to both an osmotic flow gradient and an electro-osmotic flow under the induced (Donnan) electric field.
Gathering results for the inner gradients, 
one thus deduces the expression for the average (osmotic) flow velocity as
\beq
v_w={R^2\over 8\eta} \times 2 \left[ 1-\sqrt{1+Du^2}\right] k_BT\Delta c_s
\label{vwDO}
\eeq
with $\Delta c_s=c_{\text{ext}}^R-c_{\text{ext}}^L$ defined in terms of the (reference) salt concentration in the reservoirs.

Defining the apparent osmotic pressure from $Q_w=\pi R^2 v_w= {\cal L}_w \Delta \Pi_{\rm app}$, with ${\cal L}_w={\pi R^4/ 8\eta}$, one deduces the apparent osmotic pressure $\Delta \Pi_{\rm app}$, as
\beq
\Delta \Pi_{\rm app} = 2 \left[1- \sqrt{1+Du^2}\right]  k_BT\Delta c_s
\eeq
The apparent pressure $\Delta \Pi_{\rm app}$ is of entropic origin, as highlighted by the $k_BT$ prefactor. But in a very counterintuitive way, it is predicted to be of opposite sign as the standard van 't Hoff prediction ! 
For example, for $\Delta c=c_{\text{ext}}^R-c_{\text{ext}}^L >0$, water will flow in the {\it negative} direction, {\it i.e.} towards the smallest salt concentration. This is {\it reverse} to the standard expectation for the osmotic pressure, but similar to diffusio-osmotic flows with salts. This behavior thus highlights that the apparent osmotic pressure originates in the surface contribution (even in the present thick diffuse layer regime). It is therefore attributed to  diffusio-osmosis, rather than rooted in bare osmosis. 

Now, one can introduce the rejection (or Staverman) coefficient as 
$Q_w={\cal L}_w\, \sigma(Du) \times 2 k_BT\Delta c_s$, so that
\beq
\sigma(Du)=\left[1- \sqrt{1+Du^2}\right].
\label{sigma-elec}
\eeq
It is therefore negative. Note however that this counterintuitive negative rejection is not in contradiction with the second principle, and the overall entropy production is indeed positive, as discussed in \cite{marbach2019}. As a general property,
the $L$ transport matrix in Eq.(\ref{Lmatrix}) is indeed definite positive as shown in \cite{peters2016analysis}.

We finally remind, in line with the discussion in Sec.~III.D, that the global rejection coefficient is expected to involve   contributions from various selectivity mechanisms, such as steric or dielectric rejection, acting on the transported salt across the nanochannel, see Eq.(\ref{sigma_multiple}). Accordingly the negative rejection coefficient is expected to be partly compensated by other contributions.

On the experimental side, 
negative osmotic water flow has been for example reported in the experiments by Lokesh {\it et al.} \cite{lokesh2018}, investigating water flows across functionalized carbon nanotubes (with $\sim 2.2$nm diameter) under salinity gradients. Negative fluxes were observed at low salt concentation, consistent with the rejection coefficient increasing (negatively) at low Dukhin number. 

\subsection{Alternative routes} 
\subsubsection{The integral relation}
As an alternative approach, we may consider the integral approach from equation in Eq.(\ref{eq_Piapp}), which expresses
the apparent osmotic pressure in terms of the ionic flux $j_s$. 
In the present overlapping regime of the capillary pore model, the ionic flux $j_s$ can be readily computed from from Eq.(\ref{eq2}) 
and the  Donnan expressions in Eqs.(\ref{eqDonnan2})-(\ref{eqDonnan3}), to obtain
\begin{equation}
j_s=-{2D\over L} \sqrt{1+Du^2} \left(1+\alpha_O \times (\sqrt{1+Du^2}-1)\right)\times { \Delta  c_s}
\label{fluxfinal}
\end{equation}
The first term is identical to the PNP result for $\mu_{eff}$ in Eq.(\ref{PNP-Donnan}), while the second term corresponds to 
the convective contribution to the flux under salinity gradients, introducing a dimensionless number $\alpha_O$  
\beq
\alpha_O={ {R^2\over 8\eta} 2 k_BT c_s\over D} \sim c_s R^2 \sigma
\label{alphaO}
\eeq
with $\sigma$ the ion diameter defined from the Stokes-Einstein relationship $D=k_BT/3\pi \eta \sigma$.

This dimensionless number $\alpha_O$ can be interepreted in terms of an osmotic Peclet number. The corresponding term in 
$j_s$ is of osmotic origin and corresponds to the flush of concentration under an osmotically driven velocity.
Interestingly the direction of the $\alpha_O$ term in $j_s$ suggest a flush from the high to the low concentration, hence this convective contribution to
the ion flux is rather interpreted in terms of a diffusio-osmotic contribution (in agreement with the negative rejection coefficient in this regime).

Typically, for an electrolyte confined in a nanochannel, with salt concentrations in the range $c_s\sim 10^{-3}-10^0$ M, $R=1$nm, $\sigma\sim 3\AA$, one obtains $\alpha_O \sim  10^{-4}-10^{-1}$, so the convective term would remain small compared to the diffusion one. 
However that in the presence of hydrodynamic surface slippage on the nanochannel walls (such as for carbon nanotubes), the permeability $R^2/8\eta$ is increased by a term proportional to $1+4b/R$, with $b$ a slip length. Hence the corresponding Peclet number may then largely exceeds unity. Convection cannot be neglected in such cases. 

As discussed in Sec. IV, the apparent osmotic pressure can be deduced from the (non-convective) part of the ionic flux, see
Eq.(\ref{DPi}). This corresponds to the result in Eq.(\ref{fluxfinal}) in the absence of the term proportional to $\alpha_O$. 
Using  $\Delta\Pi_{app}=k_BT\left(2\Delta c_s +{j_s}^{\rm n.c.} \times{L\over D}  \right)$, 
we deduce the same result as above, {\it i.e.}
\beq
\Delta \Pi_{\rm app} =2  \left[1- \sqrt{1+Du^2}\right] k_BT\Delta c_s
\eeq

\subsubsection{Apparent osmotic pressure from the Onsager symmetry}

From the previous calculation, we obtained that the water flow under a salt concentration gradient rewrites 
\beq
Q_w= {\cal L}_w\times \left(- \sigma(Du) c_s\right) \times -2k_BT  \Delta { \log c_s}
\label{Qosm}
\eeq 
with the negative of the ``rejection'' coefficient given by $-\sigma(Du)=\sqrt{1+Du^2}-1$ and $c_s$ is the bulk
reservoir concentration.

This result can also be infered from the symmetric transport phenomenon according to Onsager symmetry:
the excess ion flux under an (hydrostatic) pressure drop. In the latter case, the concentration and electric potential are accordingly homogeneous
inside the channel, and their values related to their external values thanks for Donnan potential. The hydrostatic pressure drops induces a flow which will flush the inner concentration of the nanopore.

Accordingly, the ion flux reduces to $j_s=c_{in}\times u=c_{in}\times {R^2\over 8\eta} \times {-\Delta P\over L}$ and
the flux in excess to the bulk concentration contribution, $j_{\text{ex}}$ writes
\beq
j_{\text{ex}}=j_s-c_{ext} v_w = (c_{in}-2c_s)\times {R^2\over 8\eta} \times {-\Delta P\over L}
\eeq
Using the value for the total salt concentration inside the nanochannels using the Donnan expression in Eq.(\ref{eqDonnan2}), one therefore obtains $c_{in}-c_{ext}=2c_s\times \left[ \sqrt{1+Du^2}-1\right]$
so that
\beq
j_{\text{ex}} = {\cal L}_w 2c_s\times \left[ \sqrt{1+Du^2}-1\right] \times (-\Delta P)
\eeq
Now, according to Onsager symmetry and Eq.(\ref{Qosm}),  the excess flux is expected to write $j_{\text{ex}} = {\cal L}_w [-\sigma(Du) \times 2c_s] \times (-\Delta P)$,
one recovers that $\sigma(Du)=1-\sqrt{1+Du^2}$, 
matching the previously obtained ``osmotic'' flow in Eq.(\ref{Qosm}).

\subsection{Osmotic ionic currents with thick layers}

We calculate here the ionic currents induced under a salt concentration gradient in the reservoirs.
From the previous PNP-S/Capillary pore expression in Eq.(\ref{eq3}), one has $i_{{e}} =  j_+-j_-= {2\Sigma \over R} v_w -e{D\over k_BT}\, c_{in} \times \left[\frac{\partial V}{\partial x}\right]_{in}$. Using the Donnan equilibrium expressions, Eqs.(\ref{eqDonnan2})-(\ref{eqDonnan3}), 
we find that 
\beq
-e{D\over k_BT}\, c_{in} \times \left[\frac{\partial V}{\partial x}\right]_{in}= -e D\times Du\times (c_{\text{ext}}^R-c_{\text{ext}}^L)
\eeq
with $c_{\text{ext}}^R-c_{\text{ext}}^L=2\Delta c_s$ and $Du={\Sigma \over R c_s}$ defined in terms of the (reference) reservoir concentration.
On the other hand, the convective contribution can be calculated using the previous result for the (diffusio-osmotic) fluid velocity, in Eq.(\ref{vwDO}),
 $v_w={R^2\over 8\eta} \times \left[ 1-\sqrt{1+Du^2}\right] 2k_BT\Delta c_s$.
 
 Altogether
 \beq
  i_{{e}} =  2{D\over k_BT}\times Du\,\beta(Du) \times c_s   {[-\Delta k_BT\log c_s]\over L}
  \label{IDOthick}
\eeq
where $\beta(Du)=   1+\alpha_O     \left( \sqrt{1+Du^2}-1\right)$, with $\alpha_O = {k_BT\over D}{R^2\over 8\eta} 2c_s$  introduced above; $\alpha_O$ fixes the order of magnitude of an osmotic Peclet number, see above. The second term thus corrects the PNP results in Eq.(\ref{PNP-Donnan}) by the corresponding convective term.

\section{Thin Double layers : diffusio-osmosis of electrolytes across nanochannels}

The thin double layer regime corresponds to the situation where the  thickness of the  diffuse layer, here the electric double layer (EDL) is  smaller than the pore size, $\lambda \ll h$. As reminded in the introduction, this is in some sense the canonical situation for diffusio-osmosis which is a surface driven phenomenon. 
The most interesting aspect is that diffusio-osmosis extends the very notion of entropically driven transport to non-selective channels, {\it i.e.} porous materials which are permeable for the solute (hence, not semi-permeable).

We obtained the expressions for the DO velocity in the sections above, see in particular Sec. III, and we briefly recap the
main results here. But beyond, I will write the various EK transport equations in the thin EDL limit, in the same spirit as for
overlapping diffuse layers. This is essential at the global scale, where transport  {\it de facto} combines the intertwinned effects
of the various transport phenomena, diffusio-osmosis but also diffusion, electro-osmosis, flushing.

We consider here a geometry of a slit of thickness $h$ but, up to geometrical factors, results apply equivalently to any other geometry. 

\subsection{Diffusio-osmotic transport, a quick recap and consequences}

\subsubsection{Solvent DO velocity and ion fluxes}

Let us assume that there is a local concentration gradient along the nanochannel, $c_0(x)$, with the concentration $c_0(x)$ defined in the middle of the channel cross section (see below).
Since $\lambda \ll h$, the flow identifies with that on a flat surface which was discussed in Sec. III. Taking a distance $z$ perpendicular to the pore surface, 
the DO velocity profile is obtained by integrating Eq.(\ref{Vx}), yielding
\begin{equation}
v_x (z)= 2 {k_BT \over \eta}  (-{\nabla_x c_s}) \times \int_0^z dz^\prime\, z^\prime\,\left[ \cosh \psi(z^\prime) -1\right] 
\label{Vxbis}
\end{equation}
with $c_s$ defined here as the salt concentration far from the surface. 
A no-slip boundary condition is assumed here, but this hypothesis is easily relaxed. The water (or solvent) velocity profile is plug-like in the nanochannel, {\it i.e.} flat beyond the 
thin EDL, with value
\beq
v_{DO}=D_{DO} (-\nabla_x \log c_s)
\eeq
where the coefficient $D_{DO}$ is defined in Eq.(\ref{DDO}) in terms of the surface potential; $D_{DO}$ has the dimension of a diffusion coefficient and typically $D_{DO}\approx { k_BT\over 4\pi \eta \ell_B} $.
As emphasized above, there is no need for semi-permeability to induce a water (solvent) flow under salinity gradient. The surface-induced diffusio-osmosis can induce such osmotic drivings. For charged solutes, the sign of the transport is however reversed as compared to the bare osmosis: the solvent flows from the high to the low salt concentration. 

Finally, and as pointed out above, the symmetry of the Onsager matrix further suggests that the same mechanisms, hence the same mobility, leads to the emergence of an excess flux generated under a pressure gradient
with~:
 \beq J_s-c_\infty Q= {{\cal A}} \times {D_{DO} \over k_BT}  (-\nabla p)
 \label{Jsexcess}
 \eeq
 with $c_\infty=2c_s$, 
 similarly to Eq.(\ref{excessflux}).

%
%

\subsubsection{Diffusio-osmotic ionic currents}

An interesting  consequence of diffusio-osmosis is the emergence of diffusio-osmotic ionic currents  
induced by a gradient of salt concentration. Again, current induced by salinty gradients is usually expected for 
selective nanochannels/membranes, such as cation- or anion- exchange membranes (CEM/AEM).
However, diffusio-osmotic mechanisms allay this selectivity constraint: channels with a pore size much larger
than the molecular scale can exhibit ionic currents under salinity gradients \cite{siria2013giant}.
In the transport matrix written in Eq.(\ref{Lmatrix}), this corresponds to the cross terms $L_{32}=L_{23}$.

Let me first make a back-of-the-enveloppe calculation to estimate the order of magnitude of the effect. Under a concentration gradient $\nabla c_s$, a 
DO velocity is generated, which in turn will carry the ions within the EDL. 
Hence the DO ionic current is expected to scale as
\beq
I_{DO} \approx {\cal P} (-e\Sigma) \times v_{DO}
\eeq
with ${\cal P}$ the perimeter of the channel (${\cal P}={2{\cal A}/ h}$ for a slit of height $h$ or $ {\cal P} =2\pi R$ for a nanotube with radius $R$); $ {v_{DO}}  =D_{DO}  \times  (-\nabla \log c_s)$ is the DO velocity as obtained above. Hence the DO current is  expected to behave like
\beq
I_{DO} \approx {\cal P} (-e\Sigma)\, D_{DO} \times \nabla \log c_{s}
\eeq
Note that this expression is actually similar to the (PNP) thick diffuse layer expression in Eq.(\ref{PNP-Donnan}) or Eq.(\ref{IDOthick}), although obtained under very different conditions.


It is however interesting to develop the full calculation of the DO ionic current (see SI of \cite{siria2013giant}).
The diffusio-osmotic current is defined as
\begin{equation}
I_{DO}= {\cal P} \int_0^{\infty}dz\, e(c_+-c_-)\, v_x(z)
\end{equation}
with $v_x$ the DO velocity profile, as obtained previously in Eq.(\ref{Vx}).
Within the PB framework, the charge density profile, $n_c=e(c_+-c_-)=2e\, c_s \sinh \psi $. 
This leads after some manipulations to
\beq
I_{DO} ={{\cal A}}\times  \mu_K  (-k_BT \nabla \log c_{s})
\label{muKDO}
\eeq
where the corresponding mobility $\mu_K$ takes the
expression
 \begin{equation}
 \mu_K=   2e{ \Sigma \over h} \times {1\over 2\pi\eta\,\ell_B}\left(1 -  {\ell_{GC}\over \lambda_D} \sinh^{-1} \left[{\lambda_D\over  \ell_{GC}}\right]\right) 
\label{mu1}
\end{equation}
with $\ell_{GC}= [2\pi\Sigma\,\ell_B]^{-1}$ the Gouy-Chapmann length and $\lambda_D$ the Debye length. 
Note that we discarded a term $\partial_x \Sigma$, assuming here a homogeneous charge along the channel. 
Finally, the diffusional contribution adds a supplementary term $I_{DO}^{\rm diff}= G \times E_{\rm diff}$, with $G$ the nanochannel conductance.

The full expression of the mobility $\mu_K$ in Eq.(\ref{mu1}) indeed reduces to the back-of-the-enveloppe estimate in the high
charge regime, {\it i.e.} $\ell_{GC}\ll \lambda_D$, for which the mobility scales as
\beq
\mu_K \underset{\text{\small large}\,\Sigma}{\propto} \Sigma.
\eeq 
 This corresponds to the regime where the PB equation is strongly non-linear. 
But in the low charge regime, {\it i.e.} $\ell_{GC}\gg \lambda_D$, one has
$1 -  {\ell_{GC}\over \lambda_D} \sinh^{-1} \left[{\lambda_D\over  \ell_{GC}}\right] \propto \Sigma^2$, 
so that the mobility scales as 
\beq
\mu_K \underset{{\Sigma\rightarrow 0}}{\propto} \Sigma^3.
\eeq 
The mobility $\mu_K$ is thus expected to be vanishingly small for small surface charges $\Sigma$. 
In other words, the DO ionic currents under salinity gradients are expected to emerge only for highly charged
surfaces, {\it i.e.} in the non-linear PB regime. 

Standard surfaces, such as glass or silica surfaces, exhibit surface charges in the tens of mC/m$^2$
and the non-linear regime,  defined as $\lambda_D> \ell_{GC}$, is only attained for salt concentrations
{\it below} $\sim 5$mM on such surfaces. Accordingly the mobility $\mu_K$ will be vanishingly small on such
surfaces for most conditions and this certainly explains why DO ionic currents have not been reported on 
such ``standard'' materials.

\begin{figure}[t!]
    \centering
    \includegraphics[width=0.5\textwidth]{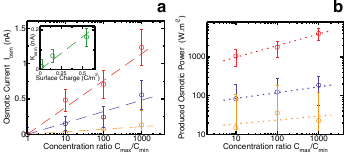}
    \caption{From Ref. \cite{siria2013giant}:  Osmotic streaming current versus concentration difference for a BNNT with
    radius and length  $\{R, L\}= \{40 {\rm nm}, 1250 {\rm nm}\}$ and pH 5.5 (yellow), 9.5 (purple) and 11 (red). 
Dashed lines are linear fits $I_{DO} = K_{osm}\times \log(C_{max}/C_{min})$. Inset, osmotic mobility versus surface charge.
Surface charge is obtained from independent conductance measurements. 
Adapted from Refs. \cite{siria2013giant} with permission from Nature Publishing, Copyright 2013.}
    \label{fig:example4}
\end{figure}

On the other hand, for a large surface charge of, say, $\sim 0.1$C/m$^2$, the non-linear regime
is reached up to salt concentrations in the molar range ($ \lambda_D > \ell_{GC}$ for $c_s <0.7$M).
Such large surface charges are more of the exception in terms of materials, but such values were indeed measured in 
specific materials such as boron-nitride nanotubes (BNNT) \cite{siria2013giant}, TiO$_2$ surface \cite{pascual2023waste}  
or activated carbon \cite{emmerich2022enhanced}. As a confirmation, strong DO ionic currents were indeed measured
on all these surfaces. As shown in Fig.\ref{fig:example4},
the experiments in Ref.\cite{siria2013giant} using BNNT fully confirmed the prediction of Eq.(\ref{mu1}): not only an ionic current is measured for non-selective nanotubes (with radius 40 nm in Fig.\ref{fig:example4}, but the current is found proportional to $\Delta \log c_s$ and the mobility is found to be proportional to the surface charge $\Sigma$ (as independently infered from conductance measurements), with a correct order of magnitude with the prediction of $K_{osm}$, see \cite{siria2013giant}.
Similar results were measured in manifold subsequent experiments
\cite{feng2016single,emmerich2022enhanced,zhang2021nanofluidics}.
Furthermore, in the experiments of Ref.\cite{emmerich2022enhanced},  the mobility $\mu_K$ on 
actived carbon surfaces (associated with high surface charges) was measured to be more than two 
orders of magnitude larger than for pristine carbon, which exhibits far smaller surface charge, hence echoing directly the above discussion.

We finally note that slippage effects leads to a strong enhancement of the mobility $\mu_K$, with an amplification
effect scaling typically like $1+b/\lambda_D$, with $b$ the hydrodynamic slip length \cite{Mouterde2019,emmerich2022enhanced}. 
However, large surface charges tend to reduce slippage via ionic friction. 

\subsection{Coupled transport in the thin layer regime}

As I discussed previously for the capillary pore model in the thick diffuse layer regime, 
the richness of the electrokinetic transport stems not only from the individual transport
mechanism but also from the coupling of the various phenomena at work: in particular 
-- in the context of this chapter --  the coupling of diffusio-osmosis with the other transport 
phenomena, diffusion, electro-osmosis and convection/flushing by liquid flow.
The global gradients imposed at the boundaries of the nanochannel usually lead to 
inhomogeneous thermodynamic fields along the channel, {\it e.g.} inhomogeneous
concentration, pressure or electric potential fields. This couples {\it de facto} the whole
EK transport mechanisms. 

The equations for a collection of pores are those introduced for the capillary-pore model
and I discuss it here in the thin diffuse layer regime. 
The model is written in full generality as a linear relationship between the local fluxes and
forces, in Eq.(\ref{Lmatrix}) \cite{gross1968membrane,peters2016analysis} and one may 
calculate directly the corresponding formulae in the thin diffuse layer limit. But I come
back here on the details of the mechanisms. 

These equations will be written here for a slit pore geometry, but this can be generalized to 
alternative, {\it e.g.} cylindrical, or more complex geometries ({\it e.g.} with varying pore size or
varying surface charge). 
In particular, such transport was also described by Poggioli {\it et al.} in the context of ionic diodes
\cite{poggioli2019};
by Cramail {\it et al.} in the context of the geometry of Surface Force Apparatus measurements \cite{cramail2025theory}, 
evidencing couplings of electrokinetic effects; mechano-sensitive effects were evidenced by Liz\'ee {\it et al.} on 
charge patterns
\cite{Lizee2025architecting}.

\subsubsection{1D-PNP equations integrating the diffuse layer contribution}

 
In the thin diffuse layer regime, there is a separation of length scales between the Electric Double Layer and the characteristic scales over which the geometry varies laterally ({\it e.g.} pore length). 
One can therefore integrate over the EDL in order to separate explicity the 
 scales and derive one-dimensional transport equations for the electrostatic potential, total ionic
concentration, pressure. 
It  echoes the approach developed in Sec. III.D for the coupling between osmosis and surface driven
diffusio-osmosis, in Eq.(\ref{Usep}). 

To simplify notation, we consider here a channel of height $h$, with  $\lambda_D \ll h$. The geometry does not really matter, as surface transport is not sensitive to boundary curvature in the thin diffuse layer limit.

The approach proceeds by separating transport along the channel length from the equilibrium properties in the EDL. 
One may directly adapt the results from the capillary model and in particular the expression of the 
ion flux and current density in Eq.(\ref{Capmodel-equ}). A direct approach following the same lines was developed in 
Ref.\cite{poggioli2019}, where, based on this formalism, it was shown that ionic diode effects could emerge even in the thin 
diffuse layer regime -- a non-trivial prediction --.

\subsubsection{Non-convective contributions}
Let me first ignore the convection effects (terms proportional to the fluid velocity), which we will estimate hereafter.
Using Eq.(\ref{Capmodel-equ}), and identifying that $c_v=c_s$ the value of the salt concentration outside the EDL, one get the cross-sectional integrals of the non-convective contributions to the solute flux and ionic density
\begin{eqnarray}
{J_{s}^{n.c.}\over {\cal A}} &= 
- {D\over k_BT}\, 2c_s \langle \cosh\psi\rangle\,\frac{\partial \mu_s}{\partial x}(x)\nonumber \\
&+ 2{eD\over k_BT} c_s \langle\sinh\psi\rangle\,\frac{\partial V}{\partial x}(x), \\
{I_e^{n.c.}\over e{\cal A}}  &= 
 {D\over k_BT}\, 2c_s \langle\sinh\psi\rangle\,\frac{\partial \mu_s}{\partial x}(x)\nonumber \\
&- {eD\over k_BT}\,2c_s \langle\cosh\psi\rangle \,\frac{\partial V}{\partial x}(x).
\label{Capmodel-equ2}
\end{eqnarray}
where the superscript $n.c.$ precise that only the non-convective contributions are calculated here.
The potential $V\equiv \phi_v$ is the electric potential in the middle of the channel. 

This involves then cross-sectional averages of  the ionic and charge distributions, $\langle c_++c_-\rangle=2c_s \langle{\cosh}\,  \psi\rangle$ and $\langle c_+-c_-\rangle = -2c_s \langle{\sinh}\,\psi\rangle$. It is easy to show that due to charge electroneutrality
$\langle c_+-c_-\rangle = {2\Sigma\over h}$ for the two interfaces. The excess concentration on a single interface can be calculated as 
\begin{eqnarray}
&{\cal F}\equiv &{\langle c_++c_--2c_s\rangle\over 2c_s} =\langle \cosh \psi \rangle -1 \nonumber \\
&& = 4 \frac{\lambda_D}{h} \left\{ \sqrt{ \left[ { \lambda_D\over \ell_{GC}} \right]^2 + 1} - 1 \right\}
\label{Fterm}
\end{eqnarray}
with again $\ell_{GC}$ the Gouy-Chapmann length. 
This term identifies to the surface contribution to the conductance. For low concentration/high surface charge, it behaves as ${\cal F}(Du)\approx  Du$,  with $Du={\Sigma\over c_s h}$ the Dukhin number.

%
\begin{eqnarray}
&{J_{\text{sol}}\over D {\cal A}} &= 2 \left[1 + {\cal F}(Du)\right]  {(-\nabla_x c_s)} +  {2eDu\, c_s  \over k_BT}{(-\nabla_x V)} ,  
\nonumber \\
&{I_e \over eD {\cal A}}&=    {2Du}  {(-\nabla_x c_s)} +{2ec_s\over k_BT}\times  \left[1+{\cal F}(Du)\right]{(-\nabla_x V)}\nonumber \\   
\label{eqn:I_general}
\end{eqnarray}
These equations identify with Eqs.(21-22) by Poggioli {\it et al.} \cite{poggioli2019} (up to a factor of 2 in the cross terms due to a difference in geometry).
Note that in these expressions the channel height $h$ and/or the surface charge $\Sigma$ may depend on $x$ as well.
Also, supplementary contributions are expected when the cation and anion diffusion coefficient are not equal.



\subsubsection{Convective effects in the thin diffuse layer regime}

Now, one should add the convective contribution, which stem from the ion transport by the solvent flow.
These are slightly more subtle to describe. 
This  follows from the  analysis  developped in the previous sections, as well as for the capillary pore model.
Gathering the previous results, the Stokes equation for the velocity field $v_w(x,z)$ along the channel writes here as
\begin{eqnarray}
0=\eta {\partial^2 v_w\over \partial z^2} &- \nabla_x \left[ p_h + 2k_BT c_s(x) (\cosh \psi -1)  \right] \nonumber \\
&- 2e c_s(x) \sinh \psi\, (-\nabla_x V)
\end{eqnarray}
where we used the expression for the pressure in Eq.(\ref{hydropv}), including the osmotic contribution, together with the
 electric driving force, $e(c_+-c_-) (-\nabla_x V)$. The longitudinal electric potential is denoted as $V$ here. 

Each of these terms has been studied in the previous sections, and one can integrate the flow profile to obtain the velocity $v_w(z)$ and
deduce its cross-sectional average
\beq
\langle v_w\rangle = {k_{hyd}\over \eta} (-\nabla_x p_h) + {D_{DO}} (-\nabla_x \log c_s) + \mu_{EO} (-\nabla_x V)
\label{vwall}
\eeq
The first term is the pressure-driven contribution with $k_{hyd}$  the hydrodynamic permeability;  the second term is the diffusio-osmotic contribution; and
the third one is the electro-osmotic contribution. The expressions for the mobility have been given in the previous sections.

Now in the presence of flow, the above results for the salt flux and ionic current in Eqs.(\ref{eqn:I_general}) also have to be corrected by the corresponding convective contributions.
The flow will flush the ionic concentrations in the EDL and this introduces supplementary terms in the expressions for the flux above
\begin{equation}
J_{\text{sol}}^{\rm conv} = {\cal A} 
\left\langle (c_++c_-)(z) {v_w(z) } \right\rangle 
\label{eqn:PNP1_nd}
\end{equation}
and
\begin{equation}
I_e^{\rm conv}= e{\cal A} 
\left\langle (c_+-c_-)(z) {v_w(z) } \right\rangle 
\label{eqn:PNP2_nd}
\end{equation}

Similarly to Eq.(\ref{vwall}), the flow profile can be written as the sum of the pressure-driven, diffusio-osmotically-driven, and electro-osmostically driven flow,
$v_w(z)=v_w^h(z)+v_w^{DO}(z)+v_w^{EO}(z)$ and each of this term will contribute to the convective flux. 
One may then calculated the various terms in the above equations, Eqs.(\ref{eqn:PNP1_nd})-(\ref{eqn:PNP2_nd}). 

However it is easier to
invoke Onsager symmetry, since these terms originating in the water flow are expected to be identical to the pressure-induced excess flux for $J_{\text{sol}}^{\rm conv}$ -- see Eq.(\ref{excessflux}) --, and to the pressure-induced streaming current for $I_e^{\rm conv}$ -- with mobility $\mu_{EO} $ --; as well as to the ionic current under concentration drop for $I_e^{\rm conv}$  -- see Eq.(\ref{muKDO}) -- and to the excess flux under the electric field -- with the  mobility $\mu_K$, so that the additional convective terms write
\begin{eqnarray}
&{J_{\text{sol}}^{\rm conv}\over {\cal A}} &-2c_s \langle v_w \rangle = {D_{DO}\over k_BT} (-\nabla p_h) + {\mu_{K}} (-\nabla_x V)
\nonumber \\
&{I_e^{\rm conv}\over {\cal A} }&=\mu_{EO} (-\nabla p_h) + \mu_{K} (- \nabla_x k_BT\log c_s)
\label{conv-contributions}
\end{eqnarray}
These terms add to the previous expressions for the solute flux and electric current, as obtained in Eq.(\ref{eqn:I_general}).
The expressions for each mobility, $D_{DO}$, $\mu_K$, $\mu_{EO}$ have been obtained in the previous sections. 


\subsubsection{Combining all elements...}

Gathering results in Eqs.(\ref{eqn:I_general}), (\ref{vwall}) \& (\ref{conv-contributions})   we have obtained the complete expression
of the local transport matrix. Differences of pressure/concentration/voltage are applied at the two ends of the channel
will induce pressure/concentration/voltage profiles inside the nanochannel, whose equations obeys the conservation laws
above, $\partial_x J_{\text{sol}}=0$ and $\partial_x I_e=0$. This is complemented with boundary condition at the channel
ends. In the limit of thin EDL, the boundary condition correspond to the continuity 
of $c_s$ and $V$ at the nanochannel borders.

The resolution of the local transport equation is by no mean obvious because of the non-linear coupling occuring between concentration
and electric field. Also when the local Dukhin number varies along the channel, {\it e.g.} under a variation of channel height or surface charge,
the solution may lead to a non-linear global response of the diode type \cite{poggioli2019}. 

In the next section, I discuss various working cases, examplifying such intertwinned transport mechanism.

\section{Working cases: when diffusio-osmosis emerges in transport across nanochannels}

We explore here various situations, inspired by experiments, where diffusio-osmotic transport couples
to other transport mechanisms to induce non-trivial emerging properties:  enhanced diffusion, mechano-sensitive
conductance and osmotic diodes.

%
%

\subsection{Example 1: Enhanced diffusion in a nanochannel}

This example is inspired by the experiments in Ref. \cite{lokesh2018}, ``{\it Osmotic Transport across Surface Functionalized Carbon Nanotube Membrane}'' by Lokesh {\it et al.}. 
Authors have studied various transport phenomena across functionalized carbon nanotubes (with $\sim 2.2$nm diameter) under salinity gradients. We previously discussed the negative water fluxes measured by the authors at low salt concentation, which are 
consistent with the negative rejection coefficient associated with Donnan rejection. 

They also report an enhanced diffusion of salt across the nanopore membrane at low salt concentration, which we discuss here.

\subsubsection{Simple estimates}

Fick diffusion is associated with a flux of salt proportional to the salt concentration gradient, as $J_s^{\rm Fick}=-D {2\Delta c_s\over L}$
(remembering that the total ion concentration is twice the salt concentration).
But observations in Ref. \cite{lokesh2018} show an enhanced salt diffusion at low concentration, see Fig.\ref{fig:example1}.
It is easy to reconcile this observation with the effect of diffusio-osmosis. According to the various descriptions above, a 
DO flow will be induced
under the concentration gradient 
obeying
\beq
v_{DO}= D_{DO} \times {-\Delta \log c_s \over L}
\label{vDO3}
\eeq
where the DO mobility has the dimension of a diffusion coefficient. 
For salt as a solute, the flow will be in the direction towards the lowest salt concentration. 

The salt flux is accordingly
\beq
J_s=-D {\Delta c_s\over L}+ v_{DO}\, 2c_s
\eeq
which using Eq.(\ref{vDO}) rewrites as 
\beq
J_s=-2(D+D_{DO}) {\Delta c_s\over L}
\eeq
In other words the DO contribution will appear as an augmented diffusion for the salt, $D_{\rm eff}=D + D_{DO}$, in agreement with the experimental results in Fig.\ref{fig:example1}. 

Now one could go further and make some estimates of the DO contribution. In Ref. \cite{lokesh2018}, the authors uses the result in Eq.(\ref{DDO}), 
$D_{DO}= { k_BT\over 2\pi \eta \ell_B} \times \log \left[(1-\gamma^2)^{-1} \right]$ with fixed surface charge, supplemented by the diffusion field contribution in Eq.(\ref{DOdiff}) for disymmetric salts. This indeed predicts an increasing effective diffusion as the salt concentration decreases.
However a large surface charge on the surface walls would be  required in the expression to obtain a (semi-)quantitative agreement.

Actually the nanotube have a small diameter (2.2 nm) and the results for thick EDL should rather apply in the present experimental configuration. In Eq.(\ref{fluxfinal}), the ionic flux is given in terms of the Dukhin number, complemented by the convective contribution
\begin{equation}
j_s=-{2D\over L} \sqrt{1+Du^2} \left(1+\alpha_O \times (\sqrt{1+Du^2}-1)\right)\times { \Delta  c_s}
\label{fluxfinal2}
\end{equation}
and we remind that $\alpha_O={ {R^2\over 8\eta} 2 k_BT c_s\over D} \sim c_s R^2 \sigma$.
The first contribution in Eq.(\ref{fluxfinal2}) is the standard diffusion modulated by the increased salt concentration inside the tube in the presence of a surface charge. 

According to Eq.({\ref{fluxfinal2}), the ionic flux is expected to strongly increase at large Dukhin number, {\it i.e.} low concentration.
However, assuming a surface charge in the tens of mC/m$^2$ range -- a reasonable value for carbon surfaces \cite{emmerich2022enhanced} --, and for a pore radius of $1$nm, the Dukin number $Du=\Sigma/c_s R$  exceeds unity only below concentrations of order $c^\star\sim 100$mM. Hence a mild increase is expected below $c^\star$. This does not really match the result in Fig.\ref{fig:example1}, where a strong increase is observed around $c_s\sim 0.4M$.
The DO contribution to the flow, expressed in the convective term proportional to $\alpha_O$, which we estimated above to be rather small with $\alpha_O\sim {\cal O}(10^{-4}-10^{-1})$.

However, as we pointed out previously, slippage effects are expected to strongly boost DO convection. In particular, $\alpha_O$, which is proportional to the 
CNT permeability, will be increased by a factor $1+4{b\over R}$ with $b$ the slip length. With a radius of 1nm, even moderate slip lengths in the tens of nanometer range will boost the DO contribution by several orders of magnitude. 
In order to go beyond and get quantitative estimates, it would be necessary to infer more specifically the surface charge inside the CNT, as well as slippage effects, and the  dependence of these parameters on salt concentration. 
However the previous argument suggests that the increase of the ionic transport results from a combination of Donnan partition, DO convection, possibly boosted by hydrodynamic slip. 

It would be interesting to reiterate exhaustive measurements in such CNT membranes in order to quantify the various transport parameters, diffusion, DO mobillity, slip length, etc., as well as the surface charge, its concentration dependence, etc.

\begin{figure}[h!]
    \centering
    \includegraphics[width=0.4\textwidth]{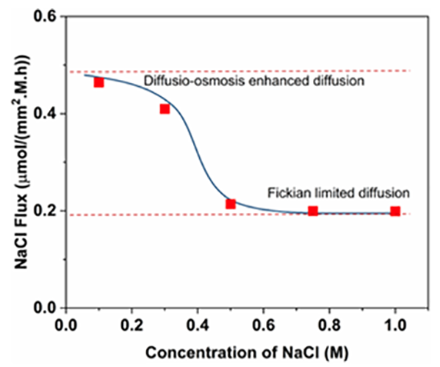}
    \caption{From Ref. \cite{lokesh2018}: enhanced diffusion across CNT membranes (b) the concentration dependence of the normalized reverse salt flux by draw solution concentration (fitting
line is for visual guidance) and (c) net osmotic water permeance under the same concentration gradient
(300 mM) generated by four different symmetric salts over salt-specific diffusio-osmotic factor:  Adapted from Ref. \cite{lokesh2018} with permission from American Chemical Society (ACS), Copyright 2018.}
    \label{fig:example1}
\end{figure}


\subsubsection{More insights into the inner transport}

The above approach gives the main trends for the effect of DO on diffusive transport. However it misses
some subtleties of the inner transport. This is what I briefly discuss here.

The diffusio-osmotic flow is expected to flush the concentration profiles, hence affecting  the concentration gradient
along the nanochannel, which will depart from the linear dependence (diffusive-like).
Accordingly the DO velocity (which is pluglike across the nanochannel section)
may depend therefore on $x$. This does not respect mass conservation (it  is not divergence free) and
 a supplementary pressure-driven flow will be built to compensate the DO flow inhomogeneities.
 Such coupling has been discussed recently in Ref. \cite{asmolov2025diffusioosmosis}, as well as in 
 Ref. \cite{lee2023role,ault2019characterization}, where the heterogeneity of the surface potential ({\it e.g.} resulting from its dependence on the local
 salt concentration) makes the description of transport particularly difficult and subtle..  
 
 Let me write the corresponding equations.
The equation for the water flux  writes
\beq
{Q_{w}\over {\cal A}}= v_{w}=-D_{DO} \times {\partial_x \log c} + {k_{hyd}\over \eta} (-\partial_x p) = {\rm cste}
\label{vtot}
\eeq
with $v_{w}$ the water  velocity (averaged over the cross section with constant area ${\cal A}$), 
$k_{hyd}$ the hydrodynamic permeability of the nanochannel and $\eta$ the fluid viscosity. 

This is complemented by the diffusion-convection equation for the total ion concentration, $c=c_++c_-$:
\beq
{J_s\over {\cal A}}=-D {\partial_x c}+ v_{w}\, c= {\rm cste}
\label{Jstot}
\eeq

Neglecting the dependence of $D_{DO}$ on salt concentration, Eq.(\ref{vtot}) integrates as
\beq
-D_{DO}\log c(x) - {k_{hyd}\over \eta} p(x) = v_{w}\, x -D_{DO}\log c(0)
\label{pprofile}
\eeq
where the boundary conditions for the total ion concentrations are $c(0)=2c_s$ and $c(L)=2(c_s+\Delta c_s)$, and
$p(0)=p(L)=0$.
One deduces
\beq
v_{w}= - {D_{DO}\over L} \log \left[ {c_s+\Delta c_s \over c_s}\right]
\label{vDO2}
\eeq
which is the same result as from the simple estimate.

On the other hand, Eq.(\ref{Jstot}) is integrated to obtain the ion flux as
\beq
{J_s\over {\cal A}}=2 {D\over L} {c_s(1-e^{{\rm Pe}_{DO}})+\Delta c_s  \over 1- e^{{\rm Pe}_{DO}}}\times {\rm Pe}_{DO}
\eeq 
with 
\beq
{\rm Pe}_{DO}= {v_{w} L\over D}
\eeq
the Peclet number defined in terms of the DO velocity in Eq.(\ref{vDO2}). The total ion concentration profile
is deduced from Eq.(\ref{Jstot}) as 
\beq
c(x)=2c_s+2\Delta c_s \times {e^{Pe_{DO}{x\over L}}-1\over e^{Pe_{DO}}-1}
\label{profilec}
\eeq
The induced pressure profile $p(x)$ is directly deduced from Eq.(\ref{pprofile}).

This shows that while the global estimate for the DO velocity in Eq.(\ref{vDO}) is captured by the simple estimate, 
the underlying DO transport leads to strong variations of the concentration profile. 

\subsubsection{Beyond: consequences on the conductance}

The detailed response under a voltage drop $\Delta V$  follows the same type of analysis. We leave
the full analysis to the reader. To my knowledge it has not been exhaustively performed. 
I rather explore here the effect of DO on the conductance, which is
the linear response to $\Delta V$: $I=G_R (-\Delta V)$

In a simple analysis, the conductance can indeed be calculated in terms of the charge carriers inside the nanochannel.
This depends on the detailed concentration profile inside the nanochannel. Assuming for simplicity that the system can be
described as a resistances in series, and neglecting surface charges for illustration, one predicts a renormalized conductance
behaving as
\beq
G_{R}=2 hw\frac{e^2 D}{k_{\rm B}T} \Bigl[\int^{L}_0  {dx\over c(x)} \Bigr]^{-1}
\label{GR}
\eeq
Using the ion concentration profile in Eq.(\ref{profilec}) predicts 
\beq
G_R=G_0 {\alpha\over 1+\alpha} \times{ \left({c_0\over c_1}\right)^\alpha - {c_1\over c_0}\over\left({c_0\over c_1}\right)^\alpha -1  }
\eeq
with $c_1=c_s+\Delta c_s$, $c_0=c_s$; $\alpha=D_{DO}/D$ and $G_0=2 hw\frac{e^2 D}{k_{\rm B}T} c_s$ the bulk conductance.
When the concentration ratio $c_1\gg c_0$, then $G_R\simeq G_0 \times D_{DO}/(D_{DO}+D)$. But in general a rather complex dependence of the conductance on the salt gradient is expected. 

The result in Eq.(\ref{GR}) predicts that the conductance is strongly affected by the DO effects, via the convective flush.
To my knowledge, the conductance under salinity gradients was not investigated exhautively experimentally. 
It would be interesting to revisit this behavior. In our own experiments in Ref.\cite{siria2013giant}, this quantity was systematically
measured but not analyzed. 


Another interesting aspect is  the effect of slippage on DO. Slippage effects are expected to boost considerably the
diffusio-osmotic mobility, as: 
$D_{DO}=D_{DO}^{\rm no-slip}\times \left( 1 + {b\over \lambda_D}\right)$ (see Sec. III.C). 
Accordingly for slip length even in the tens of nanometer range, the enhancement factor can reach hundreds,
which will considerably increase the effect of the DO flush on the conductance.

Now, the above analysis should be anyhow revisited to include surface charge contributions.
Also while we studied the concentration profile and solute fluxes, the approach should be extended 
to the analysis of local ionic current and electric potential profile.
The analysis is more involved and could possibly lead to 
ion polarization effects which are interesting to investigate.
%

\subsection{Example 2: Mechano-sensitive transport in nanochannels }

This example is inspired by the experiments of Ref.\cite{Lizee2025architecting} and their analysis,  ``{\it Architecting mechanosensitive nanofluidic transport in graphite nanoslits}'', by Liz\'ee {\it et al.}.

A number of recent experiments in nanofluidics have reported mechano-sensitive
effects, {\it i.e.} that pressure could modulate ionic transport in one way or the other
\cite{jubin2018dramatic,mouterde2019molecular,marcotte2020mechanically,paul2024nanoscale,emmerich2024nanofluidic,ismail2024mechano,paul2024mechanically}.
Such effects are intrinsically non-linear: non-linear in the global response coupling mechanical driving
(pressure) and ionic response; but then non-linear in the underlying mechanisms, which will
intertwin various transport phenomena, as those discussed above and in particular diffusio-osmosis.

As an illustration, I will discuss the foundations of such a mechanism in nanofluidic transport. 
 A recent observation made in our lab demonstrated a 
pressure-dependence of the conductance of 2D nanochannels with built-in charge patterns \cite{Lizee2025architecting}.
The geometry is shown in Fig.\ref{fig:example2}: a pressure drop (positive or negative) is applied to a nanochannel
with height of a few nanometers. The nanofabrication process allows to activate or not the surface charge at 
the entrance of the channel, in the corresponding reservoir (or inlet). If activated, the surface charge of the reservoir's surface is very large,
much larger than the charge on the channel surface.
The observation is that the contrast of charge, associated with the activation pattern, leads to a strong pressure-dependent conductance, 
 while barely any effect is measured on pristine, non-activated surface; see Fig.\ref{fig:example2}.
\begin{figure}[h!]
    \centering
    \includegraphics[width=0.5\textwidth]{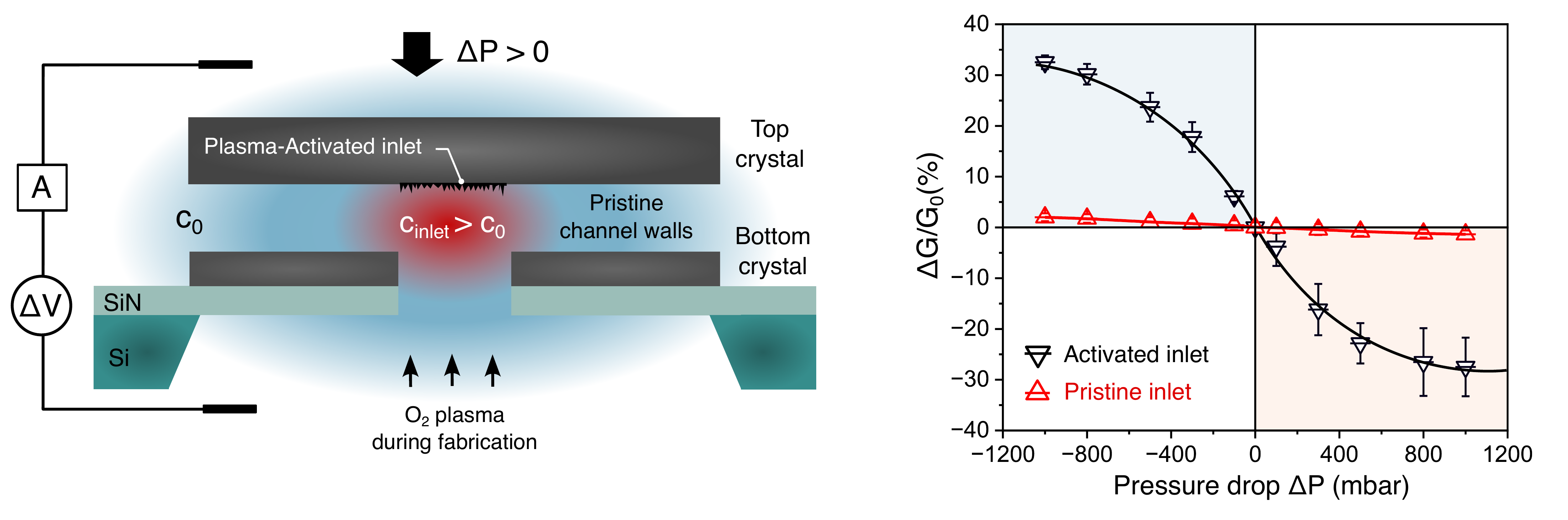}
    \caption{From Ref. \cite{Lizee2025architecting}: {Sketch of the nanofluidic experiment (left panel): a pressure drop is applied to the nanofluidic channel and the conductance is recorded. A pressure dependent conductance
    $\Delta G/G(\Delta P)$ is measured (right panel). A strong mehano-sensitive response $\Delta G(\Delta P)$ is measured for nanochannels for which the inlet surface has been activated by plasma treatement. This generates a surface charge pattern which is at the origin of the pressure dependence of the conductance.}. Adapted from Ref. \cite{Lizee2025architecting}. 
    }
    \label{fig:example2}
\end{figure}

\subsubsection{Charge contrast as a source of mechano-sensitivity}
The origin of the phenomenon is the charge discontinuity at the entrance of the nanochannel, which is at the origin of a cascade of 
transport couplings.  In a few words, the ion in the EDL close to the activated
surface are flushed towards (or away from) the nanochannel but entering the nanochannels these charges face a discontinuity of surface conductance. This leads
to an accumulation (or depletion) of ions at the entrance, which {\it in fine} changes the concentration inside the nanochannel, hence the conductance.
But this phenomenon would be negligeable without the boost of diffusio-osmosis, as we will see below.

Let me give here a minimal description for this combination of mechanisms. A more exhaustive description can be found in Ref.\cite{Lizee2025architecting}.
The starting point is the description of transport in the thin EDL limit, which I presented in Sec. VII. 
First, in the reservoir, the flow will flush the salt, including the excess ions present inside the EDL. Neglecting diffusion in this region,
one may approximate roughly   the salt flux as 
$J_{\text{sol}}\approx\langle c_++c_-\rangle v_w {\cal A}$
and assume for simplicity that ${\cal A} \sim {\cal A}_{ch}$ the nanochannel cross area, with the idea that the entrance of the nanochannel will be the limiting region for transport.
Now, we showed in  Sec. VII that the averaged concentration writes $\langle c_++c_-\rangle\simeq 2 c_s (1+{\cal F}_{R}) $, with ${\cal F}_{R}$ defined in Eq.(\ref{Fterm}) in terms of the 
activated surface charge $\Sigma_{act}$ (with a factor $1/2$ for ${\cal F}_{R}$ due to the fact that there is only one bottom surface). One deduces the ion flux originating from the flush of salt in the reservoir
\beq
J_{\text{sol}} \approx 2 c_s (1+{\cal F}_{R}) \,v_w {\cal A}.
\eeq

Now, in the nanochannel (say, $x>0$), assuming a thin diffuse layer regime, one writes a similar diffusion-convection equation for  salt transport
\beq
J_{\text{sol}} \approx 2{\cal A} D (1+ {\cal F}_{ch}) \left( \partial_x c(x) -Pe\, \frac{c(x)}{L}\right)  .
\eeq
with the P\'eclet number $Pe=v_w L/D$ and ${\cal F}_{ch}$ is defined in terms of the channel surface charge.
It is easy to obtain the general form of the solution of this differential equation.
 For low Peclet number, the expression for the salt concentration profile simplifies to 
\beq	 
c(x) = 2c_s\left(1 + Pe \, \frac{x-L}{L}\right) -\frac{J_{\text{sol}}{(x-L)}}{{\cal A}D  (1+\mathcal{F}[Du_{ch}])}  
\eeq
where we used that $J_{\text{sol}} $ is of the order of $\Pe$ and neglected  higher order terms.
Flux conservation and continuity of the concentration at the nanochannel entrance leads to
\beq
\frac{ \Delta c}{c_s}=\frac{c(0)-2c_s}{2c_s} =\Gamma \times Pe 
\eeq
where the coefficient $\Gamma$ has the expression
\beq\Gamma \approx \frac{ \mathcal{F}_R-\mathcal{F}_{ch}}{1+\mathcal{F}_{ch}} \eeq
This expression is approximate but it captures the main ingredients of the mechanism (see Ref.\cite{Lizee2025architecting}
for an extended version and comparison with the experiments).
This indeed shows that the surface charge pattern and its discontinuity leads to an accumulation or depletion of salt at the entrance
of the channel. 

This will modify the ion concentration profile inside the nanochannel, hence the conductance. The latter is defined in terms of the
number of charge carriers in the nanochannels, so that
\begin{equation}
	G= 2 hw\frac{e^2 D}{k_{\rm B}T} \left[\int^{L}_0  {dx\over c(x) + {2 \Sigma_{ch}\over h}} \right]^{-1}
\end{equation}
and after some manipulation, one predicts
\beq
{\Delta G\over G_0}={1\over 2} {1\over 1+Du_{ch}} \times \Gamma \times Pe
\eeq
with $Pe= v_wL/D$ defined as 
\beq
\mathrm{Pe} = \frac{v_w L}{D} = \frac{h^2 \Delta P}{12 \eta D} \left( 1 + 6 \frac{b}{h} \right)
\eeq
where hydrodynamic slippage is accounted for.

Altogether, this description therefore reproduces the mechanosensitive effect of the experiments, and in particular its bipolar
nature, where the conductance change is sensitive to the sign of the pressure drop. The full expression (not shown here) also
reproduces the non-linearities of the conductance in $\Delta P$ as exhibited in the experiments.

However, putting numbers, this expression is difficult to reconcile quantitatively with the experimental results, and/or only at the 
expense of unreasonable slippage effects or excessive surface charge. 

\subsubsection{The cherry on the cake: a huge diffusio-osmotic boost}

The approach above however misses an important ingredient. Indeed the created salinity gradient inside the nanochannel under
 flow is expected to induce a diffusio-osmotic velocity, $v_{DO}=-K_{DO} \times (c_s(L)-c_s(0))$ 
with  $K_{DO}=D_{DO}/c_s$. In the presence of slippage with $b\gg \lambda_D$, this yields $v_{DO}=-{k_BT\over \eta} \lambda_D b \times  (c_s(L)-c_s(0))$.
Now the concentration at the entrance varies due to the pressure itself, 
according to ${\Delta c_s/ c_s} =\Gamma\, {L\over D} v_w$. One therefore gets $v_{DO}={k_BT\over \eta} \lambda_D b \times c_s \Gamma\, {L\over D} v_w$.
%
Now, adding the bare and DO velocities, $v_{\rm eff}=v_w+v_{DO}$, one deduces
\beq v_{\rm eff}= \left( 1 + c_s{k_BT\over \eta D}\lambda_D b   \times \Gamma \right) \times v_w
\eeq
We therefore obtain an strongly enhanced effective Peclet number, renormalized by the diffusio-osmotic transport:
\beq 
Pe_{\rm eff} = Pe \times \left( 1 + c_s{k_BT\over \eta D}\lambda b  \times \Gamma \right) \gg Pe
\eeq

In other words, the DO transport strongly amplifies the flow effects on the salt concentration profile, which itself
is the origin of the DO flow. Slippage plays a key role in this amplification loop. 
But it is only at the expense of the inclusion of all the coupled transport effects that a consistent
qualitative and quantitative description of the transport can be achieved. 
This is an interesting
lesson, which is at the core of the richness of nanofluidic transport. 

\subsection{Example 3: Beyond van 't Hoff, osmotic diodes and its applications for desalination}
This example is inspired by the theoretical work of Ref.\cite{picallo2013nanofluidic} ``{\it Nanofluidic Osmotic Diodes: Theory and Molecular Dynamics Simulations}'' and the experiments of Ref.\cite{abdelghani2025resonant}, ``{\it Resonant osmotic diodes for voltage-induced
water filtration across composite membranes}''.

A seminal step in the short history of nanofluidics is the demonstration of ionic diodes  \cite{karnik2007rectification,cheng2010nanofluidic}.
Based on an analogy to PN junctions in semi-conducting materials, nanochannels with patterned surface charges were shown to induce a non-linear current-voltage response, in a form very similar to the Schokley diode. The analogy is quite deep, in the sense that the transport equation for anions and cation in nanofluidic channels -- the PNP equations described above -- is analogous to those for electrons and holes in semiconducting system, with the surface charge playing the role of the doping. Ionic diodes were shown in manifold conditions and various geometries: patterned nanofluidic systems, conical nanopores, nanopipettes, etc. Rectification occurs in asymmetrics nanochannel, as quantified by a gradient of the Dukhin number $\nabla \left[{\Sigma\over c_s g}\right]$. This shows incidently that Debye layer overlap is actually not a prerequise to obtain current rectification, as shown in \cite{poggioli2019}.

Now, a key difference between semi-conducting systems and electrolytes is that electrolytes are liquid which flow under stress, while semi-conducting systems are solid. One therefore expects that the rectification of the ionic current in asymmetric channels should extend to flow, and in particular osmotically driven flows.
Interestingly, rectified osmosis with asymmetric water permeability under solute gradient reversal was actually reported in a number of biological cell systems, 
such as eyhtrocytes \cite{farmer1970perturbation}, granulocytes \cite{toupin1989permeability}, epithelial cells \cite{chara2005asymmetry}, or COS-7 fibroblasts \cite{peckys2011rectification}.

%
%
%
%


A simple geometry to investigate the diode effect for osmotic flows is to consider a nanochannel with an 
asymmetric surface charge as sketched in Fig.\ref{fig:example3}: the left side has a positive surface charge density $e\Sigma$ while the right side has a negative surface charge $-\alpha e\Sigma$, with $\alpha > 0$ a numerical coefficient and $\Sigma$ a surface density.
We impose a salt concentration $c_L$ on the left side and $c_R$ on the right side, as well as 
an additional voltage drop $\Delta V = V_R-V_L$. 
One defines the Dukhin number as $Du_0=\Sigma/hc_s$, with $c_s={1\over 2}(c_L+c_R)$ the averaged salt concentration. 
Following Ref.\cite{picallo2013nanofluidic},
several assumptions allow developping a full analytical calculation:  (i) the nanochannel is thin enough so that the thick diffuse layer calculations apply, see  Sec. VI;   (ii) convective effects are neglected, {\it i.e.} the  P\'eclet number is small; (iii) the Dukhin number is assumed to be larger than unity (this approximation is not a prerequisite but simplifies calculations).
 The calculation then extends the simplified PNP description in Sec. VI.A to the composite geometry considered here. Care has to be
 taken of the boundary conditions at each charge discontinuity, at the two ends of the channel and at the middle interface: as discussed in Sec. VI.A,
 there is a discontinuity of ion concentrations and electric potential due to the Donnan potential contributions.
 
 We refer the reader to the supplementary materials of Ref.\cite{picallo2013nanofluidic} for the complete calculation along these lines. This yields the final result for the solute flux under the concentration and voltage drop:
\begin{equation}\label{eq:js}
j_s ={D\over L}\left(2\frac{{c}_L-{c}_R}{Du_0 }-\frac{\alpha-1}{\alpha\, Du_0 } {{c}_R\over c_0}\left[{c}_L e^{{e\Delta{V}\over k_BT}} -{c}_R \right]\right)
\end{equation}
%
Now, in order to calculate the apparent osmotic pressure, we will make use of the integral relationship between $\Delta\Pi_{\rm app}$ and the ionic flux $j_s$ obtained in Eq.\eqref{eq_Piapp}: $\Delta\Pi_{\rm app}=k_BT\left(2\Delta c_{ s} +{j_s}\times{L\over D}  \right)$. 

This leads to the expression of the apparent osmotic pressure in the high surface charge regime, as
\begin{eqnarray}
\label{eq_Piapp2}
&{\Delta \Pi_{\text{app}} }=&2k_BT\left(1-{1\over Du_0}\right)(c_R-c_L)\nonumber \\
&&- 2 k_BT {\cal C} \left[ e^{{e\Delta{V}\over k_BT}} -{{c}_R \over c_L}\right] 
\end{eqnarray}
where ${\cal C}=c_R c_L(1-\alpha)/(\alpha Du_0 (cR+c_L))$ is a non-linear function of concentrations, and has the dimension of a concentration.
This  apparent osmotic pressure generates a water flow as
\begin{equation}
Q_w= {{\cal A}\over L} {\cal L}_{\rm hyd}\, {\Delta \Pi_{\text{app}} }[c_R,c_L,\Delta V]
\label{Qtotdiode}
\end{equation}

In these expressions, the driving force is indeed proportional to $k_BT$ and one may conclude that it is therefore of entropic entropic  (it does not simply reduce to an electro-osmotic effect). Furthermore the flow is not necessarily directed towards the highest concentration, as for bare osmosis. The sign of the non-linear contribution to $\Delta \Pi_{\text{app}}$ (or equivalently $Q_w$)
depends on the sign of the surface charge contrast $\propto (\alpha -1)$ (as well as on $\Delta V$). This is a surface osmotic effect and 
reflects a non-linear diffusio-osmotic transport. 
\begin{figure}[h!]
    \centering
    \includegraphics[width=0.5\textwidth]{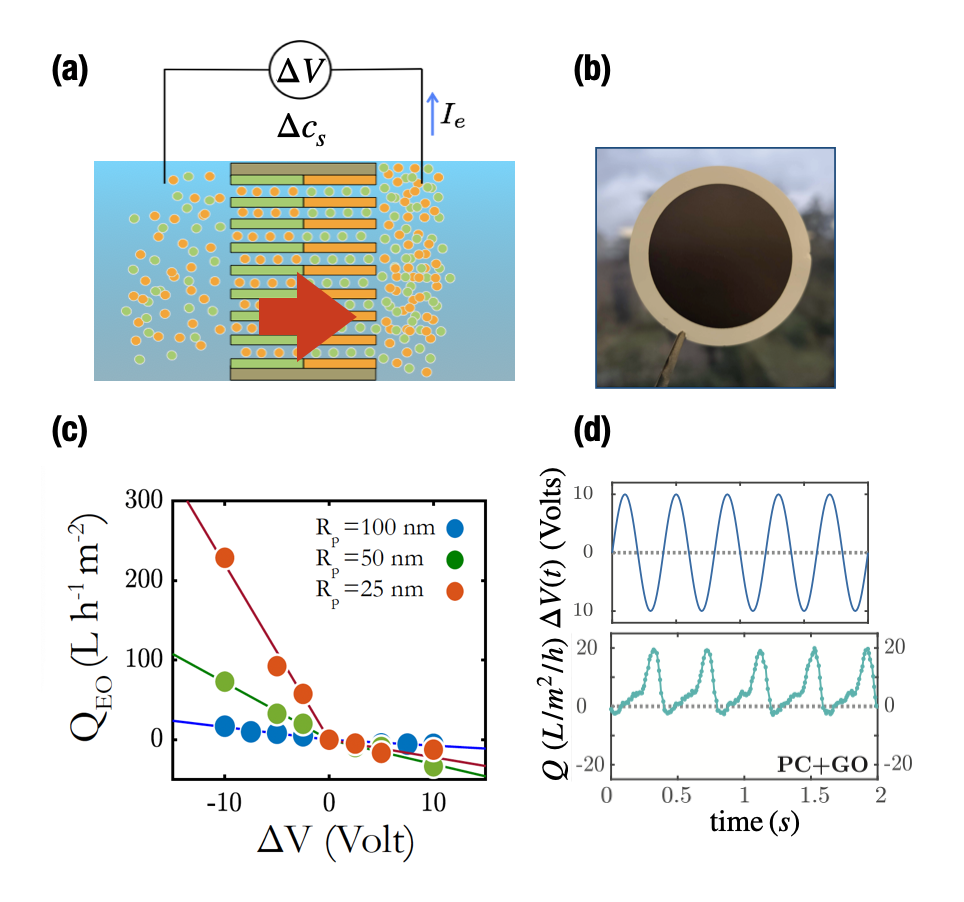}
    \caption{Flow rectification with composite membranes: (a) Sketch of a composite membranes with asymmetric nanopores; (b) Example of a realization using a mesoporous polycarbonate (PC) membrane on which a graphene oxide microporous membrane is deposited; (c) Flow under (DC) voltage drop showing voltage rectification; (d) A sinusoidal (AC) electric driving yields a rectified water flow: the averaged flow is non-vanishing even if the driving force averages to zero.
     From Ref. \cite{abdelghani2025resonant}:  Adapted from Refs. \cite{abdelghani2025resonant} with permission from Springer-Nature, Copyright 2025.}
    \label{fig:example3}
\end{figure}

%
%


The ``osmotic'' diode, which one may coin a van 't Hoff diode, shows rectification both in the concentration drop and in the applied voltage drop. 
Such rectification effects have been recently observed at the membrane scale in Ref.\cite{abdelghani2025resonant}. This work makes
use of an asymmetric assembly ot two membranes which are welded together: one mesoporous layer with large pores (in the tens of nanometer
range) and a surface charge, while the second microporous layer has small pores dedicated to induce rejection of a targeted solutes (its surface charge being different from the other side). In the experiments of Ref.\cite{abdelghani2025resonant}, a porous polycarbonate membranes
was used for the large pore side, while the selective part was of various types, such as graphene oxide, MoS$_2$, or reverse osmosis membranes (by Toray). In simple terms, water flow is driven within
the mesoporous layer, while selectivity is achieved within the microporous
layer. These asymmetric membranes do exhibit a rectified flux under a voltage driving, see the flux versus the applied voltage in Fig.\ref{fig:example3}. Furthermore under a sinusoidal electric driving, a net water flux will be induced by the rectification. 
Finally, an interesting way to quantify the performance of the electric is to measure how much pressure is equivalent to a given voltage  
for a targeted flux. This conversion factor $\alpha$, measured in bars per Volts per bar, is found in the experiments to reach as much as 15 bars per volt for appropriate materials ! Meaning that 2 applied volts being equivalent to 30 bars.

%
This development paves the way for the practical implementation of ionic diodes as a viable filtration technology. In this design, the ``electric'' driving mechanism substitutes the conventional pressure-driven process, thereby eliminating the need for the high-pressure pumps typically employed in most filtration systems, especially in reverse osmosis. Moreover, the flow rectification feature enables the use of a sinusoidal voltage drop, which significantly reduces polarization effects and generates a localized, oscillating flush at the membrane surface -- an effective means of mitigating fouling. This concept has since been scaled up by the spin-off company Ilion Water \cite{Ilion}, which now manufactures reverse osmosis vessels based on this technology.

\subsection{Example 4: The diffusio-osmotic perspective for osmotic energy harvesting }


It is not my purpose here to develop an exhaustive discussion of osmotic energy harvesting and I will only make a general introduction to the concept. Over the recent years, there has been a surge of exploration of osmotic transport across low-dimensional materials -- boron-nitride and carbon nanotubes and, more recently, two-dimensional (2D) membranes such as graphene, hexagonal boron nitride, transition-metal dichalcogenides, pristine and activated graphite, etc. --, 
which constitute  new platforms for the study of osmotic energy conversion. The literature has become considerable and I refer the reader to exisiting reviews which have flourished recently, see in particular \cite{post2007salinity,logan2012membrane,siria2017new,macha20192d,zhang2021nanofluidics,abidin2022towards}.

This discussion of this topics  is however interesting in the context of this chapter since the introduction of diffusio-osmotic transport for osmotic energy harvesting has helped renewing the concept, which was somewhat in a technological dead-end. 

 I discovered the question of blue energy in the context of our fundamental work ``{\it Giant osmotic energy conversion measured in a
single transmembrane boron nitride nanotube}'' in Ref.\cite{siria2013giant}, where we showed that diffusio-osmosis-based mechanisms could induce large power densities for some specific materials, while relaxing some of the usual constraints of osmotic harvesting. This work has been the start of a personal journey, from the lab to the creation of a company, {\it Sweetch Energy} \cite{Sweetch}, which has  developped a scaled up osmotic generator based on the nanofluidic principles, now implemented in an industrial pilot plant. 

\subsubsection{Osmotic energy, a quick recap}
Osmotic, or ``blue'', energy denotes the free energy that is released when solutions of different salinity mix, for example where rivers discharge into the sea. This mixing process is associated with a substantial entropy gain, corresponding to an ideal extractable energy of about 0.8 kWh per cubic metre of freshwater mixed with seawater. This is a quite old idea which goes back to a suggestion by Pattle in 1954 \cite{pattle1954production} and later by Normal and others in the 70's \cite{norman1974water}.

The global potential for electricity generation is estimated in the range of a few thousands of TWh per year, depending on the assumptions and technologies used. Going more into the detailed potential based on the world hydrology, the installed capacity would be of order 250-300 GW with an extraction factor of 20\% (similar to the total installed nuclear power worldwide). This is considerable and by no mean anecdotical. Key advantages of osmotic power are its non-intermittent character, it is dispatchable, widespread, and dense in terms of industrial facility per unit surface (a back of the enveloppe calculation for the required installation using equivalent numbers for existing desalination plants would suggest  up to $\sim 10^2$W installed per m$^2$ of ground, far above solar and wind power figures).
Finally, keeping in mind that the cost of installation of 1GW of nuclear power is typically in the 6 to 10 billions\$ (CAPEX), it seems worth digging a bit more on osmotic energy and its potential.

Classical osmotic power technologies rely on membrane-based separation processes operating between a high-salinity (seawater or brine) and a low-salinity (river or treated wastewater) reservoir. Two architectures have been predominantly explored at the pilot scale: pressure-retarded osmosis (PRO) and reverse electrodialysis (RED). In PRO, a water-permeable, salt-rejecting membrane converts the osmotic pressure difference (up to ~30 bar at a seawater-river interface) into hydraulic work on a turbine. 
In RED, stacks of cation- and anion-selective membranes spatially separate counter-ion fluxes driven by the concentration gradient, directly generating an electrical current, yet practical power densities also remain typically below a few watts per square metre of membrane. Despite successful demonstration projects, including PRO and RED plants in Norway, Japan and the Netherlands, membrane modules using classical technologies still fall short of the $\sim5$W/m$^2$ threshold generally considered necessary for economic viability under realistic pretreatment and pumping costs.  

The origin of this low performance is easy to understand. For example in PRO, although the osmotic pressure at the seawater-freshwater interface is high, reaching about 30 bar, the semi-permeable membrane has an extremely low permeability because its pores are on the sub-nanometre scale to exclude ions. As a result, the volumetric flow rate remains very small, and since power is given by the product of flow rate and pressure drop, the resulting power output is correspondingly limited.
This efficiency bottleneck is rooted in nanoscale transport phenomena inside the membranes, most notably the trade-off between selectivity and permeability imposed by molecular sieving and ion exchange. 

Nanofluidic principle, with the emergence of sometimes exotic transport may help bypass these limitations. This has been actually the route followed by the company Sweetch Energy \cite{Sweetch} which took inspiration from nanofluidic principles to develop  efficient, cheap, large scale osmotic modules, bypassing the $\sim5$W/m$^2$ osmotic threshold.


\subsubsection{Diffusio-osmosis as an alternative lever for osmotic energy harvesting}

The concept of diffusio-osmosis has changed the perspective for  osmotic energy conversion  by showing that it does not require strict molecular selectivity as previously conceived and can instead be driven by interfacially controlled mechanisms such as diffusio-osmosis. 
This directly echoes (and clarifies) my statement in the introduction: diffusio-osmotic transport extends the conditions of applicability of 
standard osmosis to less constrained conditions. As I exhaustively discussed in this chapter, diffusio-osmotic transport allows to express osmotic drivings in channels and membranes without the prerequisite of semi-permeability and high rejection, so that osmotic transport can occur even in non-selective pores. 
This is a considerable asset, with important consequences.

Leaving the rhetoric aside,  let me remind a few key relevant results.  
From the previous sections, see Sec.VII.C, electrical currents are generated under chemical gradients 
via diffusio-osmosis according to 
\beq
I_{DO} = K_{osm} \Delta \log c_{s}, 
\label{IDO-V2}
\eeq
where $K_{osm}={{{\pi R^2}\over L}}\times  \mu_K  k_BT$ and $\mu_K$ is the corresponding mobility, see Eq.(\ref{mu1}) for the thin diffuse layer limit and Eq.(\ref{IDOthick}) for the thick diffuse
layer. For large surface charge, the thin diffuse layer result is
\begin{equation}
 \mu_K \simeq   2e{ \Sigma \over R} \times {1\over 2\pi\eta\,\ell_B}
\label{mu1bis}
\end{equation}
As discussed above, the thin diffuse layer result demonstrates the generation of ionic currents across
non-selective nanochannels. 

As we discussed in Sec.VII.C., 
such DO ionic currents  were measured in manifold  experiments
\cite{siria2013giant,feng2016single,emmerich2022enhanced,zhang2021nanofluidics}.
In  \cite{siria2013giant}, both the dependence of the ionic current as $\nabla \log c_s$ and
the linear dependence of $\mu_K$ with the surface charge $\Sigma$ has been experimentally verified, see Fig.\ref{fig:example4}.

Also, as we discussed in the previous sections, the expression for the mobility $\mu_K$ in Eq.(\ref{mu1}) is recovered only  
for highly charged surfaces, 
while $\mu_K$
is vanishingly small for less electrified surfaces. The measurement of diffusio-osmotic currents across BN nanotubes
in \cite{siria2013giant} was then a serendipitous observation since BN has since proved to be a highly charged materials,
with the surface charge increasing for smaller radius BN nanotubes. The surface electrification of BN is singular by many aspects
and make it an interesting topics to investigate \cite{grosjean2016chemisorption,grosjean2019versatile,cetindag2023anomalous,wang2025spontaneous}.

Such studies does not restrict to osmotic energy harvested from salinity gradients. It could be also represent an interesting lead to 
harvest waste heat and in particular the so-called low grade heat, with temperatures between 25 and 100$^\circ$C. This is a huge and untapped source, from power generation and several industrial sectors. The energy waste was estimated  in the range of 2500TWh per year for Europe only \cite{luberti2022estimate},  to compare with the primary energy source of 13000 TWh per year. It is however challenging to valorize this reservoir. A proposal is to use liquid-based osmotic energy conversion in a closed loop. It relies on a liquid-liquid phase separation harvested the low grade heat, and the mixture is then recombined at low temperature in a osmotic energy converter to harvest the corresponding (entropic) free energy \cite{logan2012membrane}. In \cite{pascual2023waste}, phase separating ionic liquids were used
and the energy harvesting during recombination used diffusio-osmotic energy conversion on a titanium-dioxyde porous membrane. 
Powers densities in the range of $\sim 7$W/m$^2$ were measured on TiO$_2$ membranes, making such approaches promising at large scale.


%


\subsubsection{Scaling up nanofluidics }
Exploiting such surface-driven transport in highly charged nanopores actually provides a route to circumvent the conventional permeability--selectivity compromise and to boost osmotic power densities as compared with PRO and RED. Interestingly, while existing PRO and RED plant are highly optimized in terms of engineering, the core of the process, the membranes, still rely on pre-existing designs developped for desalination or other applications. 
This is the main contribution of nanofluidics, which -- via exquisiste studies under super controlled conditions -- allow identifying new prescriptors for the development of optimal membrane design dedicated to osmotic energy harvesting, in particular in terms of materials, porosity, surface chemistry. 
Such principles allow circumventing the main challenges, in particular charge polarization, but also electric resistance in the reservoirs (low salinity), inlet resistances, etc. This strongly allays the conclusions of \cite{wang2021nanopore}.

Using such ``nanofluidic-informed'' principles to conceive new membranes and stack design, the individual osmotic module of the company Sweetch Energy (with $\sim$ 20m$^2$ membrane) -- this is the lego brick of an osmotic plant -- has bypassed the $\sim5$W/m$^2$ osmotic ``break-even'', allowing the company to install an industrial pilot plant on the Rh\^one river \cite{Sweetch}. 

Now, going beyond, there are many possible developments still ahead and
I will only cite a few (subjectively chosen) leads. At the laboratory scale, nanochannels have demonstrated record  power densities ranging from kilowatts to megawatts per square metre of effective active area, owing to the combination of molecular-scale thickness and extremely high surface charge densities.
These results indicate that properly engineered nanoporous 2D membranes could, in principle, overcome the limitations of conventional polymer and ceramic membranes and make osmotic energy conversion competitive at scale. Boron-nitride emerged as a material of choice for osmotic energy harvesting \cite{siria2013giant,li2024ion} and its performance have been scaled up at the membrane scale \cite{casanova2020enhanced,pendse2021highly,cetindag2023anomalous}. But scaling up BN and similar materials at the scale of hundred of thousands of square meter -- which is the required scale for an osmotic plant -- is quite hopeless, both technologically and financially.
Beyond such materials, key direction of the recent research is to explore and screen various materials to optimize osmotic energy conversion, such as MOF, cellulose,  silk-based materials, MXenes, etc. \cite{zhang2021nanofluidics,jiang2025mxene,li2018hybrid,xin2019high}. Some works have also tackled more fundamental aspects affecting osmotic harvesting process, such as the concentration and charge polarization, which strongly limitates the performance, see {\it e.g.}  \cite{sripriya2024nanofluidic},  the use of ionic diodes \cite{siria2017new,zhang2015engineered,li2018hybrid}, or capacitive  instead of faradic electrodes \cite{chapuis2025boosting}. 
 
Realizing the potential of osmotic power, however, requires bridging a substantial gap between single-pore physics and membrane-scale devices. Challenges include scalable fabrication of cheap, large-area  membranes with high pore density, preserving mechanical robustness and fouling resistance in complex feed waters, furthermore assembled in an osmotic cell designed to minimize the global electric resistance.
This is a huge challenge, embraced by very few actors. 
At the same time, advances in surface chemistry, controlled doping and heterostructure design open possibilities for tailoring charge density, wetting and even nonlinear ionic transport, for example through ionic diode effects in asymmetric nanopores, to further enhance energy extraction. In this context, osmotic energy sits at the intersection of materials science, surface chemistry and nanofluidics. This is a domain where  the emerging understanding of nanofluidic transport at the smallest scales can really make a splash for innovation.

\section{Conclusions and perspectives }


This chapter has become excessively long and I explored the principles of diffusio-osmosis in (far too much) detail. But in doing so, 
I was naturally led to a broader discussion of ion and fluid transport in nanochannels. Taken together, these electrokinetic transport mechanisms form a single, tightly coupled framework, within which the role of diffusio-osmosis turns out to be far more central than one might initially assume.
I  hope that this chapter will equip readers with the necessary tools to analyse the rich variety of exotic behaviours that emerge in nanofluidic systems.
My hope is that the tools developed here -- from mechanical force balances to capillary-pore and PNP-Stokes descriptions, from thin to overlapping double layers, and from local mobilities to global Onsager matrices -- will help readers dissect and interpret the growing zoo of `exotic' behaviours observed in nanofluidics, in line with those that I briefly discussed: negative rejections, mechano-sensitive conductance, osmotic diodes, giant blue-energy currents, and others still to come.

At the same time, the present chapter barely scratches the surface of what diffusio-osmosis could do when coupled to other ingredients. Many directions remain open and there are many topics that I barely discussed, in spite of their considerable interest:
\begin{itemize}[leftmargin=2pt]
\item[-] diffusio-osmosis shows up whenever there is a gradient of a solute close to a surface. Obviously, the list of gradient sources is immense as discussed in the context of diffusio-phoresis in Ref.\cite{velegol2016origins}. Gradients naturally occurs in chemistry -- with chemical reactions at work --, membranes and filtration -- with the occurence of fouling or concentration and charge polarization \cite{cho2016non}--, in biology -- with the ubiquituous gradients in cellular compartments--, but also in geology --with reactive nanopores \cite{plumper2017fluid,roman2025inhibition} and the importance of ultra-slow phenomena on geological time scale--, etc.
\item[-] in most of the chapter, I focused for simplicity on symmetric salts, with no effects of the asymmetry of diffusivity. This is a question of interest since diffusivity asymmetry will induce supplementary diffusion electric fields, inducing the dynamical equivalent to a ionic Hoffmeister serie, and make the analysis and coupling even more subtle \cite{abecassis2008boosting,henrique2022impact}.
\item[-] most of the descriptions reported in this chapter involve continuum description while the phenomena at stake occurs at nano- and even sub-nano- scales. Molecular effects and deviations from continuum are expected to play a decisive role, which remain to fully rationalize. In particular, even the molecular definition of the microscopic stress tensor raises fundamental questions \cite{liu2018pressure,liu2017microscopic} and this is a topics which desserves further investigations.
\item[-] transient effects and time-dependent osmosis, in line with the notion of frequency dependent osmotic pressure \cite{marbach2020resonant,marbach2017active} which remain to be generalized to diffusio-osmotic transport.
\item[-] I skipped here a discussion on diffusio-phoretic transport since it is {\it a priori} barely related to nanochannels, the core of the present chapter. But the topics is considerable, both in terms of fundamentals and applications \cite{marbach2019}. It relates also to active matter, where chemical gradients can power artificial swimmers. Using gradients as sources of energy also apply to the development of nano-engines, and diffusio-phoretic and -osmotic processes have been harnessed  to power nanorods to rotate \cite{shi2022sustained}, for example mimicking biological rotary molecular engines such as ATP synthase. 
\item[-] the discussion on diffusio-osmosis immediately generalize to thermo-osmosis, whose  level of subtlety even exceeds diffusio-osmotic mechanisms \cite{ganti2017molecular,herrero2022fast}
\item[-] finally, a topics of great interest is the mechanical deformation of the solid surface under diffusio-osmosis. Diffusio-osmosis is related to osmotic pressure gradients and {\it de facto} this induces forces, as discussed for phoresis in \cite{marbach2020local,mckenzie2022drop} and explored recently for diffusio-osmotic transport \cite{bonthuis2014mechanosensitive,maroundik2025diffusioosmotic}. In Ref.\cite{bonthuis2014mechanosensitive}, it was shown with finite-element calculations of the PNP equations that a deformable nanochannel may transition between a closed and open states on the basis of the diffusio-osmotic stress caused by the ions inside the channel. This is a fascinating perspective.
\end{itemize}

Beyond the academic appeal, these questions are not just intellectual curiosities. Diffusio-osmosis is already informing the design of high-efficiency blue-energy membranes, and voltage-driven filtration concepts, with first industrial-scale implementations now on the horizon. 
If the reader is convinced of the considerable potential of diffusio-osmosis in nanochannels, this chapter will have achieved its main goal.

\section*{Acknowledgements}
I thank all my colleagues from the n-AQUA network for many interactions, as well as support from ERC project n-AQUA, grant agreement $101071937$. 
I thank Daan Frenkel for nice discussions during the writing of this chapter.

\section*{Appendix: Reminder on Poisson-Boltzmann (PB) solution}
The solution of the PB equation
\beq
{d^2\over dz^2} \psi=-{1\over \lambda_D^2} \sinh \psi
\eeq
(with $ \psi = eV/k_BT$ the dimensionless potential) can be obtained as
\beq
\psi = -2 \log\left( {1+\gamma e^{-z/\lambda_D}\over 1-\gamma e^{-z/\lambda_D}}\right)
\eeq
where the coefficient $\gamma$ is the solution of the equation
\beq
\gamma^2 + 2 {\ell_{GC}\over \lambda_D} \gamma -1 = 0
\eeq
with $\ell_{GC}$ is the Gouy-Chapman length ($\ell_{GC}\sim 1/\vert \Sigma\vert$). 

One has
\beq
\gamma=\sqrt{1+\left({\ell_{GC}\over \lambda_D}\right)^2}-{\ell_{GC}\over \lambda_D}
\eeq

Further details on the PB solution can be found in \cite{andelman1995electrostatic} and \cite {herrero2024poisson}.
%
%
%

\bibliographystyle{rsc}
\bibliography{references} 

\end{document}